\documentclass[a4paper,11pt]{article}
\pdfoutput=1 

\usepackage{jcappub} 
\usepackage{xspace}
\usepackage{xcolor}
\usepackage{booktabs}
\usepackage{multirow}
\usepackage[noabbrev]{cleveref}
\usepackage{lineno}
\usepackage{hyperref}
\usepackage{orcidlink}
\usepackage{amsmath}
\usepackage{amsfonts}
\newcommand{\lya}{Ly$\alpha$\xspace}
\newcommand{\lyaf}{Ly$\alpha$ forest\xspace}
\newcommand{\ion}[3]{#1\thinspace{}#2(#3)\xspace}

\newcommand{\lyaxlya}{Ly$\alpha\times$Ly$\alpha$\xspace}
\newcommand{\combinedLBG}{Ly$\alpha\times$Ly$\alpha+$Ly$\alpha\times$LBG\xspace}

\newcommand{\lyaxlbg}{Ly$\alpha\times$LBG\xspace}
\newcommand{\lyaxlae}{Ly$\alpha\times$LAE\xspace}

\newcommand{\kms}{\mathrm{km\,s^{-1}}}
\newcommand{\Mpc}{\mathrm{Mpc}}
\newcommand{\hMpc}{h^{-1}\,\mathrm{Mpc}}
\newcommand{\hoverMpc}{h\,\mathrm{Mpc}^{-1}}
\newcommand{\ap}{\alpha_\parallel}
\newcommand{\at}{\alpha_\perp}
\newcommand{\aiso}{\alpha_{\rm ISO}}
\newcommand{\aAP}{\alpha_{\rm AP}}
\newcommand{\rp}{r_\parallel}
\newcommand{\rt}{r_\perp}

\usepackage{fontawesome5}
\usepackage{physics}

\makeatletter
\newcommand{\github}[1]{%
\href{#1}{\faGithub}\footnote{\url{#1}}%
}
\makeatother

\DeclareMathOperator{\sinc}{sinc}

\defcitealias{DESI.DR2.BAO.lya}{\texttt{DESIDR2-Ly$\alpha$}}
\newcommand{\DESIDRIILya}{\citetalias{DESI.DR2.BAO.lya}}

\defcitealias{2023RamirezPerez:LyaCatalogEDR}{\texttt{RP2023}}

\defcitealias{Ruhlmann-Kleider:2024LBG}{\texttt{R24}}
\newcommand{\RuhlmannLBG}{\citetalias{Ruhlmann-Kleider:2024LBG}}

\defcitealias{Payerne:2024LBGs}{\texttt{P2024}}

\title{\boldmath The Lyman-$\alpha$ Forest from LBGs: First 3D Correlation Measurement with DESI and Prospects for Cosmology}
\emailAdd{herreraa@iap.fr}
\affiliation{Affiliations are in Appendix \ref{sec:affiliations}}

\author[1,2]{{Hiram K. Herrera-Alcantar}\orcidlink{0000-0002-9136-9609},}
\author[2]{{Eric~Armengaud}\orcidlink{0000-0001-7600-5148},}
\author[2]{{Christophe~Yèche}\orcidlink{0000-0001-5146-8533},}
\author[3]{{Calum~Gordon}\orcidlink{0000-0003-2561-5733},}
\author[3]{{Laura~Casas},}
\author[3]{{Andreu~Font-Ribera}\orcidlink{0000-0002-3033-7312},}
\author[2]{{Christophe~Magneville},}
\author[4]{{Corentin~Ravoux}\orcidlink{0000-0002-3500-6635},}
\author[6]{{J.~Aguilar},}
\author[7]{{S.~Ahlen}\orcidlink{0000-0001-6098-7247},}
\author[6]{{A.~Anand}\orcidlink{0000-0003-2923-1585},}
\author[3]{{D.~Brooks},}
\author[6]{{E.~Chaussidon}\orcidlink{0000-0001-8996-4874},}
\author[6]{{T.~Claybaugh},}
\author[6]{{A.~Cuceu}\orcidlink{0000-0002-2169-0595},}
\author[8]{{K.~S.~Dawson}\orcidlink{0000-0002-0553-3805},}
\author[9]{{A.~de la Macorra}\orcidlink{0000-0002-1769-1640},}
\author[10]{{Arjun~Dey}\orcidlink{0000-0002-4928-4003},}
\author[3]{{P.~Doel},}
\author[6,11]{{S.~Ferraro}\orcidlink{0000-0003-4992-7854},}
\author[12,13]{{J.~E.~Forero-Romero}\orcidlink{0000-0002-2890-3725},}
\author[14,15,16]{{E.~Gaztañaga}\orcidlink{0000-0001-9632-0815},}
\author[6,17]{{S.~Gontcho A Gontcho}\orcidlink{0000-0003-3142-233X},}
\author[18]{{A.~X.~Gonzalez-Morales}\orcidlink{0000-0003-4089-6924},}
\author[19]{{G.~Gutierrez},}
\author[6]{{J.~Guy}\orcidlink{0000-0001-9822-6793},}
\author[20]{{C.~Hahn}\orcidlink{0000-0003-1197-0902},}
\author[21]{{D.~Kirkby}\orcidlink{0000-0002-8828-5463},}
\author[6]{{A.~Kremin}\orcidlink{0000-0001-6356-7424},}
\author[3]{{O.~Lahav},}
\author[6]{{A.~Lambert},}
\author[6]{{M.~Landriau}\orcidlink{0000-0003-1838-8528},}
\author[22]{{L.~Le~Guillou}\orcidlink{0000-0001-7178-8868},}
\author[23,4]{{M.~Manera}\orcidlink{0000-0003-4962-8934},}
\author[24,25,26]{{P.~Martini}\orcidlink{0000-0002-4279-4182},}
\author[10]{{A.~Meisner}\orcidlink{0000-0002-1125-7384},}
\author[27,4]{{R.~Miquel},}
\author[9]{{A.~Muñoz-Gutiérrez},}
\author[15]{{S.~Nadathur}\orcidlink{0000-0001-9070-3102},}
\author[2,6]{{N.~Palanque-Delabrouille}\orcidlink{0000-0003-3188-784X},}
\author[28,29,30]{{W.~J.~Percival}\orcidlink{0000-0002-0644-5727},}
\author[31]{{F.~Prada}\orcidlink{0000-0001-7145-8674},}
\author[32]{{I.~P\'erez-R\`afols}\orcidlink{0000-0001-6979-0125},}
\author[33]{{G.~Rossi},}
\author[34]{{E.~Sanchez}\orcidlink{0000-0002-9646-8198},}
\author[6]{{D.~Schlegel},}
\author[35,36]{{M.~Schubnell},}
\author[6]{{J.~Silber}\orcidlink{0000-0002-3461-0320},}
\author[10]{{D.~Sprayberry},}
\author[36]{{G.~Tarl\'{e}}\orcidlink{0000-0003-1704-0781},}
\author[10]{{B.~A.~Weaver},}
\author[6]{{R.~Zhou}\orcidlink{0000-0001-5381-4372},}
\author[37]{{H.~Zou}\orcidlink{0000-0002-6684-3997},}

\usepackage{arydshln}
\usepackage{enumitem}

\date{January 2025}
\abstract{The Lyman-$\alpha$ (\lya) forest is a key tracer of large-scale structure at redshifts $z > 2$, traditionally studied using the spectra of luminous but relatively rare quasars. In this work, we explore the viability of using the fainter yet significantly more abundant Lyman Break Galaxies (LBGs) as alternative background sources for \lya forest studies. We analyze 4,151 \lya forest skewers extracted from LBG spectra obtained in the DESI pilot surveys conducted in the COSMOS and XMM-LSS fields. From this dataset, we present the first measurement of the \lya\ forest auto-correlation function derived exclusively from LBG spectra, probing comoving separations up to 48~$\hMpc$ at an effective redshift of $z_{\rm eff} = 2.70$. The measured LBG \lya\ forest auto-correlation is consistent with that derived from DESI DR2 quasar \lya\ forest spectra at a comparable redshift, validating the use of LBGs as reliable background sources for \lya\ forest analyses. In addition, we measure the cross-correlation between the LBG \lya\ forest and the positions of 13,362 galaxies, demonstrating that this observable serves as a sensitive diagnostic for assessing the precision and accuracy of galaxy redshift estimates, and for identifying and correcting systematic offsets. Finally, using both synthetic LBG spectra and Fisher matrix forecasts, we show that a future wide-area survey covering $\sim$5,000~$\deg^2$, targeting 1,000 LBGs per square degree at signal-to-noise levels comparable to our sample, could enable LBG-based \lya forest baryon acoustic oscillation (BAO) measurements with expected uncertainties of $\sigma_{\aiso} = 0.4\%$ (isotropic) and $\sigma_{\aAP} = 1.3\%$ (Alcock–Paczynski). This performance is further enhanced when combining the BAO analysis with a \lya\ forest Full Shape (FS) approach, yielding a predicted uncertainty of $\sigma^{\mathrm{FS}}_{\aAP} = 0.6\%$. These results open a new avenue for precision cosmology at high redshift using the \lya\ forest in dense LBG samples.}

\begin{document}
\maketitle
\flushbottom
\section{Introduction}\label{sec:introduction}
The large-scale statistical properties of the matter distribution encode a wealth of information about our Universe, tracing its expansion history, primordial fluctuations and non-Gaussianities, and the growth of cosmic structure and the formation of galaxies. For that reason, a number of surveys such as DESI~\cite{2019BAAS...51g..57L,DESI2016b.Instr}, EUCLID~\cite{Euclid:2021icp}, PFS~\cite{2014PASJ...66R...1T}, and WEAVE~\cite{2012SPIE.8446E..0PD} are ongoing or planned to measure these cosmological structures over a wide range of scales and redshifts. At redshifts $2 < z \lesssim 4$, a remarkably powerful probe is the Lyman-$\alpha$ (\lya) forest. The \lyaf is a set of absorption features imprinted onto the optical spectra of background sources. These arise from resonant Lyman-$\alpha$ absorption and scattering due to the presence of neutral hydrogen along the line-of-sight. Given that most of this \lyaf signal comes from regions located in the diffuse intergalactic medium (IGM)~\cite{2009RvMP...81.1405M}, these spectroscopic features encode information on the physical properties of this medium~\cite{2019ApJ...872...13W}.

At the same time, the \lyaf can be considered as a biased tracer of the large-scale matter distribution. The 3D correlations between absorption features from different lines-of-sight, first measured in~\cite{Slosar:2011LyaxLya}, can be interpreted using a linear clustering model, valid on large, linear scales, and incorporating relevant astrophysical and instrumental contaminants. As a consequence, the baryon acoustic oscillation (BAO) feature at a typical redshift of $z \sim 2.3$ could be measured from the large \lyaf samples collected by the BOSS and eBOSS surveys~\cite{2013A&A...552A..96B,2013JCAP...04..026S,2020ApJ...901..153D} and more recently by the DESI survey~\cite{2025JCAP...01..124A,DESI.DR2.BAO.lya}. With the highest precision achieved on the isotropic BAO distance scale delivered by DESI data release 2 (DR2) results (\cite{DESI.DR2.BAO.lya}, hereafter \DESIDRIILya) at the 0.7~\% level. It should be noted that these BAO measurements combine two correlation signals: the auto-correlation of the \lyaf, and the cross-correlation between \lya absorption and the spatial distribution of galaxy tracers, as pioneered in~\cite{Font-Ribera:2013LyaxQSO}.

Until now, most of the background sources used to carry out \lyaf observations were quasars (also denoted as QSOs). Indeed quasars are among the brightest optical sources in the Universe, with a strong continuum flux in the rest-frame UV wavelength range. Therefore spectra with a good signal-to-noise ratio can be measured from those sources up to high redshift, in relatively short exposure times with existing telescopes. However quasars are rare objects in our Universe, especially when restricting to those with redshift $z > 2.1$, which have their \lyaf in the optical observed wavelength range. The quasar luminosity function was measured over a wide range of redshift and magnitude from SDSS spectroscopic observations in~\cite{2006AJ....131.2766R,2016A&A...587A..41P}: predictions from models fitted in~\cite{2016A&A...587A..41P} indicate that, in the redshift range $2<z<3$, there are only $\sim 140$ quasars per deg$^2$ with magnitude $r<24$, and down to the deeper limit of $g<25$ the expected number count is $\sim 230$ per deg$^2$. In practice, the DESI survey obtained spectra for 60 high-redshift  ($z>2.1$) quasars per deg$^2$, using a photometric selection~\cite{2023ApJ...944..107C} limited to $r<23$. Such a sky density, while unprecedented over a wide footprint, remains a limiting factor for several scientific applications. For instance, it restricts measurements of line-of-sight absorption cross-correlations between close quasar pairs~\cite{2003MNRAS.341.1279R,2024JCAP...05..088A}, and also affects the detection of the BAO feature, which remains in the shot-noise-limited regime for DESI~\cite{DESI2016a.Science}.

In order to significantly increase the sky density of \lyaf samples, background sources other than quasars are needed. Lyman Break Galaxies (LBGs) are prime candidates, as these star-forming galaxies are among the most readily detected at high redshifts. Their distinctive break at the Lyman limit (912~\AA) enables efficient selection from broadband photometry~\cite{Ruhlmann-Kleider:2024LBG}. However, LBGs are typically fainter than quasars, so extracting the \lyaf signal from their spectra requires longer exposure times, larger telescopes, or both. Despite these challenges, the \lyaf in high signal-to-noise ratio LBG spectra has already been exploited in relatively small-area surveys such as CLAMATO~\cite{Hess:2018qit,Horowitz:2021zwh} and LATIS~\cite{Newman:2020iao}. While the primary goal of these programs was to map the three-dimensional distribution of the IGM at high redshift and to identify structures in the cosmic web (e.g., protoclusters, voids, filaments), they also enabled measurements of the Ly$\alpha$–galaxy cross-correlation at $z>2$ on scales $r<30\ \hMpc$ by combining forests extracted from both quasars and LBGs~\citep[e.g.,][]{Momose2021,Momose2021LAE,Newman2024,Zhang2025}. These efforts complement works probing the Ly$\alpha$-galaxy connection based exclusively on quasar forests~\citep[e.g.,][]{Adelberger2003,Adelberger2005,Bielby2017, Liang2021LAE}.

In this article, we extend these efforts by demonstrating the potential of using the \lyaf in LBG spectra alone as an independent tracer of large-scale structure in the context of wide-area spectroscopic surveys. We use optical spectra of high-redshift galaxies collected during DESI pilot observations in the COSMOS and XMM–LSS fields, as described in \Cref{sec:dataset}. In \Cref{sec:analysis,sec:results}, we present the first 3D \lyaf auto-correlation measurement using only LBG forests, covering scales from 1 to 48 $\hMpc$, along with the \lya cross-correlation with both LBGs and Lyman-$\alpha$ emitters (LAEs). We show that these measurements can be described with the same modeling framework applied to DESI \lyaf correlations from quasars. Finally, in \Cref{sec:prospects}, we provide forecasts for high-redshift BAO measurements, highlighting that future wide-area surveys with DESI-like capabilities can exploit the \lyaf in LBG spectra to deliver competitive cosmological constraints at redshifts beyond those accessible with quasars alone.

\section{Dataset}\label{sec:dataset}
DESI is a fully robotic, highly multiplexed spectroscopic instrument installed on the 4‑meter Mayall Telescope at Kitt Peak National Observatory \citep{DESI2022.KP1.Instr}. It delivers simultaneous spectra for up to 5,000 targets over a 3.2$^\circ$ diameter field of view~\citep{DESI2016b.Instr,FocalPlane.Silber.2023,Corrector.Miller.2023,FiberSystem.Poppett.2024}. The majority of DESI’s dark‑time observing program~\citep{SurveyOps.Schlafly.2023} is dedicated to its main survey, which aims to acquire spectra for roughly 50 million galaxies, including several million quasars at $z\gtrsim1.6$. In addition, DESI has conducted dedicated observing campaigns in the COSMOS and XMM–LSS (hereafter XMM) fields during both the survey validation and main-survey phases~\citep{Joshua:inprep}, including targeted pilot programs to test and refine observation strategies for high-redshift galaxies such as Lyman-break galaxies (LBGs) and Lyman-$\alpha$ emitters (LAEs).

In the following section, we describe the target selection strategies and the automated redshift estimation pipeline used to construct the LBG and LAE samples for this work. Preliminary analyses of the \lyaxlbg and \lyaxlae cross-correlations revealed systematic redshift offsets of $\Delta v_{\rm LBG} = -169 \pm 22\ \kms$ for LBGs and $\Delta v_{\rm LAE} = -241 \pm 20\ \kms$ for LAEs. To mitigate potential biases in the clustering signal, such as a shift in the  cross-correlation along the line-of-sight due to these offsets, we apply a correction to the redshift estimates in all subsequent measurements. Details of these corrections are provided in Appendix \ref{appendix:redshift_correction}.

\subsection{Target Selection}\label{sec:target_selection}
LBG spectra exhibit a pronounced flux ``break” just blueward of the 912 \AA\ Lyman limit in the rest frame, primarily caused by absorption from neutral hydrogen in stellar atmospheres and the interstellar medium (ISM). At high redshifts ($z > 2$), additional \lya\ forest absorption caused by the intergalactic medium, further suppresses the flux blueward of 1216 \AA. These characteristic features enable efficient identification of Lyman-break galaxies (LBGs) at $z \sim 2.3$ via broad-band photometry, particularly through dropout selection in the $u$-band.

Our target selection employs the $u$-dropout technique, based on deep imaging from the Hyper Suprime-Cam survey~\citep[HSC;][]{HSC19}, complemented by $u$-band imaging from the Canada-France-Hawaii Telescope’s CFHT Large Area U-band Deep Survey~\citep[CLAUDS;][]{Sawicki19}, obtained with MegaCam.

We adopt two complementary $u$-dropout selection strategies. The first, described in~\cite{Ruhlmann-Kleider:2024LBG} (hereafter \RuhlmannLBG), applies color cuts in the $u-g$ vs $g-r$ plane, alongside a magnitude threshold in the $r$ band. The second approach, presented in~\cite{Payerne:2024LBGs}, utilizes a random forest algorithm trained on HSC+CLAUDS $ugriz$ photometry artificially degraded to the depth of the Ultraviolet Near-Infrared Optical Northern Survey~\citep[UNIONS;][]{2025UNIONS}, forecasting selection performance for upcoming surveys such as DESI-II.

A smaller subset of our sample is selected using medium-band photometry. In this case, Lyman-alpha emitters (LAEs) are identified by flux excesses in narrow-band filters, while LBGs are selected based on flux decrements between adjacent bands. We emphasize that throughout this work, the term ``LAE" \emph{refers exclusively to the photometric selection technique}, rather than to the strength of the \lya\ emission.

All LBG and LAE targets presented here were observed in DESI pilot programs conducted between 2021 and 2024 in the COSMOS and XMM fields, with effective exposure times ranging from 2 to 5 hours. In our \lya\ forest analysis (see \Cref{sec:analysis}), only the LBGs serve as background sources. The median $r$-band magnitude of the LBG sample is $r=23.94$.

\subsection{Automated Redshift Measurement}\label{subsec:redshift_measurements}
Spectra of the LBG and LAE targets were obtained with the DESI 3-arm spectrographs, which cover the wavelength range 3,600--9,800 \AA\ with a spectral resolution between 2,000 and 5,000~\citep{DESI2022.KP1.Instr}. Given the redshift range of these sources, most of the key spectral features fall within the blue arm (3,600--5,930 \AA), where the resolution varies from 2,000 to 3,000. Raw spectra were processed using the DESI spectroscopic pipeline~\citep{Spectro.Pipeline.Guy.2023}, and multiple exposures of the same object were coadded with weights proportional to their signal-to-noise ratio (SNR). The vast majority of the spectra were obtained using a dithering strategy, in which each target was typically observed multiple times with different fibers (and, in some cases, different spectrographs) across overlapping tiles following a rosette pattern. This configuration averages out potential systematics from sky-line subtraction residuals and minimizes fiber and spectrograph-dependent effects, which reduces possible biases in the clustering measurements.

Redshift estimation proceeds in two stages. First, a convolutional neural network (CNN) identifies whether the spectrum corresponds to an LBG or LAE and assigns a preliminary redshift. Second, a template-fitting algorithm refines the redshift determination.

The CNN is a modified version of \texttt{QuasarNET}~\citep{Busca18,Green:2025nvv}, which DESI uses for quasar classification. For LBG classification, we retrained the CNN using 14 interstellar absorption lines and 2 emission lines (see Table~3 in \RuhlmannLBG). Each of the 16 lines are assigned a CNN confidence level ($\mathrm{CL}$) and are sorted in descending order. A spectrum is classified as an LBG or LAE if its fifth-highest $\mathrm{CL}$ exceeds a chosen threshold. Additional details on the CNN algorithm and performance can be found in Section 4.1 of \RuhlmannLBG.

To refine spectral classification and the redshift estimate, the redshift of the line with the highest CNN confidence is used as a prior for the \texttt{redrock} software~\github{https://github.com/desihub/redrock}~\citep{Redrock.Bailey.2024}, DESI’s standard classifier and redshift-fitting tool. We employ a customized version of \texttt{redrock} that includes four LBG templates derived from visually confirmed spectra in \RuhlmannLBG: one template without \lya\ emission and three with increasing \lya\ emission strength.

To quantify the quality of the redshift measurements of our COSMOS LBG sample, we cross-matched our dataset with the high-confidence ($\mathrm{CL} > 0.95$) spectroscopic redshifts from the COSMOS spectroscopic redshift compilation~\citep{Khostovan2025}. We define catastrophic failures as sources with a velocity offset $|\Delta v| > 3,000~\kms$ relative to the reference spectroscopic redshift. The resulting outlier rates are 16.5\%, 13.3\%, 11.1\%, and 8.3\% for thresholds of $\mathrm{CL}>0.97, 0.99, 0.995,$ and $0.999$, respectively. These values are consistent with those reported in Table 4 of \RuhlmannLBG\ and confirm that higher CL cuts substantially reduce the catastrophic outlier rate, although a non-negligible fraction remains even at our strictest cut. As will be noted in \Cref{subsec:param_contraints}, sample purity and catastrophic redshift outliers directly impact the inferred clustering amplitude on the correlation functions. Therefore, it will be an important consideration for future large-scale clustering analyses with LBG forests. We return to this point in \Cref{sec:summary_conclusions}.

For the non-catastrophic objects, we find excellent agreement with the reference catalog, with a median velocity offset of only $\sim 8~\kms$ and a scatter of $\sigma_{\rm MAD} = 1.4826 \times \mathrm{MAD}(|\Delta v|) \sim 180~\kms$. This demonstrates that, aside from the catastrophic failures, the redshifts of our CNN+$\texttt{redrock}$ pipeline are robust and suitable for clustering analyses.

\begin{figure}
    \centering
    \includegraphics[width=\textwidth]{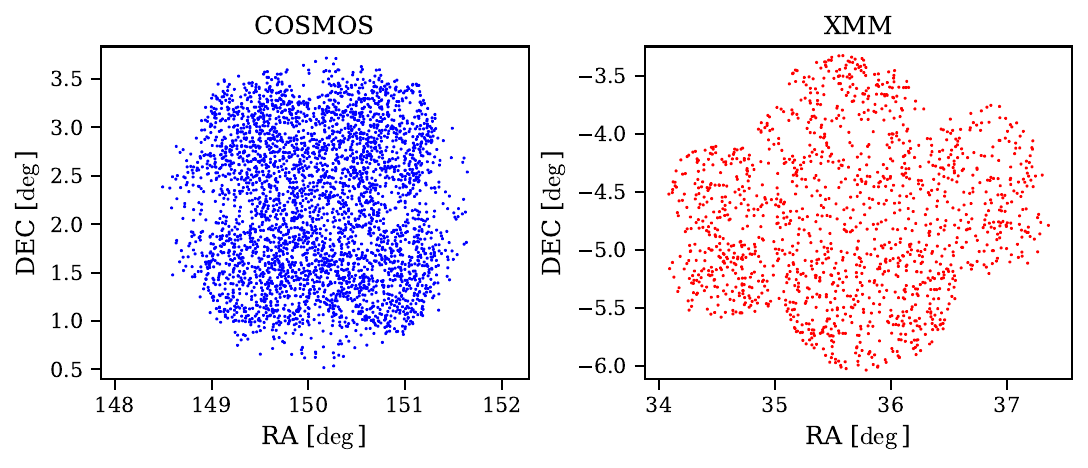}
    \caption{Angular positions of the $\textrm{CL}>0.995$ LBG sample in the COSMOS (left) and XMM (right) fields used for our analysis. The COSMOS footprint covers approximately $6~\deg^2$, while the XMM footprint spans about $5~\deg^2$.}
    \label{fig:footprint}
\end{figure}

For our baseline analysis, we adopt a confidence threshold of $\mathrm{CL}>0.995$ to enhance sample purity while preserving statistical power. We also consider thresholds in the range $0.97 < \mathrm{CL} < 0.999$ for robustness checks. \Cref{tab:statistics} summarizes the number of LBGs, LAEs, and accepted \lya forests (see \Cref{sec:analysis}) in COSMOS and XMM as a function of $\mathrm{CL}$ threshold. At $\mathrm{CL}>0.995$, the sample comprises 4,931 LBGs, 8,431 LAEs, and 4,151 forests; the angular distribution of the LBG targets is illustrated in \Cref{fig:footprint}. The survey footprints correspond to approximately $6$ and $5\ \deg^2$ in COSMOS and XMM, respectively.

We note that approximately 70\% of LBGs in our sample are in COSMOS, while 70\% of LAEs are in XMM. This imbalance in the field distributions is a direct result of the strategy employed by the DESI pilot survey campaigns. Different targeting strategies were tested in each field; specifically, the COSMOS field observations were optimized for LBG selection to enable future Ly$\alpha$ tomography studies, while LAEs were targeted mainly in XMM where medium-band imaging was available first. Future DESI special programs are planned to observe both fields with more uniform strategies, which will provide larger and more balanced samples for future analyses.

\begin{table}[htp]
  \centering
  \caption{Statistics of LAEs, LBGs, and selected \lya forest skewers for different $\mathrm{CL} $ cuts.}
  \label{tab:statistics}
\resizebox{\textwidth}{!}{%
  \begin{tabular}{llrrr}
    CNN Confidence Level& Survey & LBG tracers & LAE tracers & \lya Forests (\Cref{sec:analysis}) \\    \toprule
    \multirow{2}{*}{$\mathrm{CL}>0.97$}
    & COSMOS         & 4,847 &  3,495 & 4,117 \\
    & XMM            & 1,988 &  7,100 & 1,555 \\
    \hline
    \addlinespace
    \multirow{2}{*}{$\mathrm{CL}>0.99$}
    &COSMOS         & 3,951 &  2,733 & 3,395 \\
    &XMM            & 1,598 &  6,430 & 1,258 \\
    \hline
    \addlinespace
    \multirow{2}{*}{$\mathrm{CL}>0.995$}
    & COSMOS         & 3,517 &  2,416 & 3,027 \\
    & XMM            & 1,414 &  6,015 & 1,124 \\
    \hline
    \addlinespace
    \multirow{2}{*}{$\mathrm{CL}>0.999$}
    & COSMOS         & 2,714 &  1,848 & 2,336 \\
    & XMM            & 1,080 &  4,969 &  859 \\
  \end{tabular}}
\end{table}

\section{Data Analysis}\label{sec:analysis}
One of the main goals of this work is to present the first measurement of the \lyaf\ auto‐correlation (\lyaxlya) extracted solely from LBG spectra, together with the corresponding cross‐correlations between \lya\ absorption and LBG (\lyaxlbg) and LAE (\lyaxlae) positions. Our analysis builds on the \DESIDRIILya\ framework for quasar‐forest studies, implemented via the public \texttt{picca} package~\github{https://github.com/igmhub/picca}~\citep{picca}.

Below, we outline the steps of our correlation‐function measurement pipeline, highlighting the minor differences with respect to the quasar-based pipeline. For further methodological details, we refer the reader to the \DESIDRIILya\ reference and to Section 3 of \cite{2023RamirezPerez:LyaCatalogEDR}.

\subsection{Continuum Fitting and Extraction of Transmitted Flux Field}\label{subsec:delta_extraction}
The first step in our analysis is to preprocess the spectra. Following \DESIDRIILya, this normally comprises a series of operations but we do not apply all of them here. We mask bad pixels identically to \DESIDRIILya, removing regions affected by galactic absorption, residual sky-line emission, and CCD defects flagged by the DESI reduction pipeline~\citep{Spectro.Pipeline.Guy.2023} (e.g., cosmic ray contamination or hardware-related issues). On the other hand, we do not recalibrate the fluxes as the redshift range ($2.0 < z < 3.5$) of our LBG sample is not wide enough to completely cover a calibration region (typically C\thinspace III; see Table 3 of \cite{2023RamirezPerez:LyaCatalogEDR}) in the 3,600--5,423~\AA\ observed frame wavelength range that we have adopted for our analysis.

Unlike quasar forest analyses, we do not mask Damped Ly$\alpha$ Absorbers (DLAs) or Broad Absorption Line (BAL) features. BALs are quasar-specific phenomena and are not expected to appear in LBG spectra. DLAs, however, are present in LBG spectra and are expected to impact our measurements. Unfortunately, no dedicated DLA catalog exists for our LBG sample. Existing DLA detection algorithms, developed for quasar spectra, cannot be straightforwardly applied here due to the differing SNR characteristics and different spectral shapes LBGs with respect to quasars. Furthermore, DLAs in low-SNR quasar spectra (SNR $< 3$) are typically not masked in quasar forest analyses (e.g., \DESIDRIILya), and most of our LBG spectra fall below this SNR threshold. Adapting any of the existing quasar‐based DLA finders~\citep[e.g.,][]{Wang:2022CNNDlaFinder,Ho:2020DLAFinderGP,Y3.lya-s2.Brodzeller.2025} is beyond the scope of this work.

After spectral preprocessing, we extract the transmitted‐flux field,
\begin{equation}
	\delta_g(\lambda) = \frac{f_g(\lambda)}{\overline{F}(\lambda)C_g(\lambda)} - 1,
	\label{eq:deltaflux}
\end{equation}
where $g$ indexes each LBG, $f_g(\lambda)$ is the observed flux, $\overline{F}(\lambda)$ is the mean transmission, and $C_g(\lambda)$ is the unabsorbed LBG rest-frame spectrum, hereafter referred to as the continuum. We limit our analysis to the $1040-1205$\AA\ rest-frame wavelength interval and the $3600 <\lambda_{\rm obs}< 5423$ \AA\ observed wavelength range, which corresponds to a redshift range of $1.98 < z < 3.5$.

As in quasar analyses, we model $\overline{F}(\lambda)C_g(\lambda)$ by a universal rest‐frame continuum, $\overline{C}_{\rm LBG}(\lambda_{\rm RF})$, modified by a first-degree polynomial in $\Lambda = \log \lambda$ with two free parameters, $a_g$ and $b_g$,
\begin{equation}
	\overline{F}(\lambda)\  C_g(\lambda) = \overline{C}_{\rm LBG}(\lambda_{\rm RF}) \left(a_g + b_g\ \frac{\Lambda - \Lambda_{\rm min}}{\Lambda_{\rm max} - \Lambda_{\rm min}}\right).
\end{equation}
We determine $\overline{C}_{\rm LBG}$, $a_g$, and $b_g$ through an iterative likelihood maximization process, simultaneously fitting the pixel‐variance,
\begin{equation}\label{eq:flux_variance}
	\sigma_g^2 = \eta(\lambda)\  \sigma^2_{{\rm pipe},g}(\lambda) + \sigma^2_{\rm LSS}(\lambda) \left[\overline{F}(\lambda) C_g(\lambda)\right]^2,
\end{equation}
where $\sigma^2_{{\rm pipe},g}(\lambda)$ is the reported variance from the data processing pipeline~\citep{Spectro.Pipeline.Guy.2023}, $\eta(\lambda)$ corrects for small inaccuracies in the noise estimate,\footnote{The \DESIDRIILya\ pipeline treats $\eta(\lambda)$ as a free parameter and found values close to unity with small fluctuations. Given the limited size of our dataset, we decided to fix $\eta(\lambda) = 1$.} and $\sigma^2_{\rm LSS}(\lambda)$ is a free parameter that captures the intrinsic \lya flux variance. Because our LBG spectra is dominated by noise, we were unable to robustly constrain $\sigma_{\rm LSS}$ in our analysis. We nevertheless include this term to maintain a consistent methodology with the DESI \lya\ quasar analyses and verified that omitting it leads to negligible differences in the measured correlation functions. Future, larger LBG spectroscopic samples should enable a more precise determination of $\sigma_{\rm LSS}$.

\begin{figure}[!htbp]
    \centering
    \includegraphics[width=\textwidth]{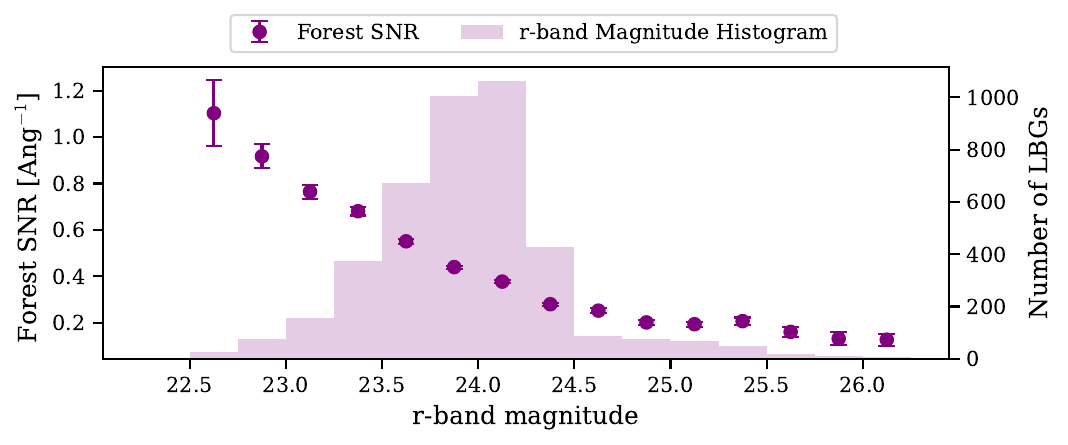}
    \caption{Filled histogram: $r$-band magnitude distribution of LBGs with $\mathrm{CL}>0.995$ in the COSMOS and XMM fields, whose \lya forests pass our quality cuts. The mean forest SNR as a function of $r$-band magnitude is shown with purple points; error bars represent the standard error of the mean. The effective exposure times for these observations range from 2 to 5 hours, with a median value of 4.5 hours. A small number of bright LBGs have longer exposures due to repeated observations.}

    \label{fig:mag_vs_snr}
\end{figure}

Additionally, we apply quality cuts to remove forest skewers with (i) negative fitted continuum, (ii) negative mean SNR in the forest region, or (iii) fewer than 150 valid pixels. This yields 3,027 accepted forests in COSMOS and 1,124 in XMM. The median SNR of spectra in the forest region is 0.35 \AA$^{-1}$, which is about 4.5 times lower than the median SNR of 1.57~\AA$^{-1}$ found in the quasar \lya\ forests sample used in \DESIDRIILya. The distribution of SNR as a function of $r$-band magnitude is shown in \Cref{fig:mag_vs_snr}. 

\begin{figure}[!htbp]
    \centering
    \includegraphics[width=\textwidth]{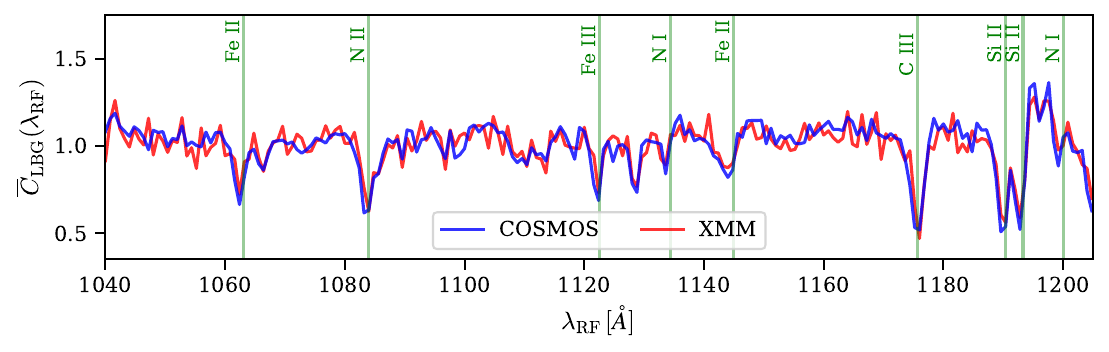}
    \caption{Mean rest-frame continuum in the \lya region as measured from LBG spectra on the COSMOS (blue line) and XMM (red line) fields. The green shaded vertical lines indicate known spectral line features caused by stellar and interstellar absorption. By construction from the continuum fitting procedure, $\overline{C}_{\rm LBG}$ is arbitrarily normalized.}
    \label{fig:mean_flux}
\end{figure}

\Cref{fig:mean_flux} shows the mean rest-frame continuum $\overline{C}_{\rm LBG}$ in both fields, revealing excellent agreement between COSMOS and XMM. Moreover, both samples exhibit the characteriztic absorption features from stellar and interstellar lines, consistent with those identified in composite LBG spectra from previous studies~\citep[e.g.,][]{Shapley:2003LBG,Newman:2020iao,Monzon:2020knq,Ruhlmann-Kleider:2024LBG}. Notably, the continuum-fitting procedure as implemented in \texttt{picca}, originally developed for quasar \lya forest analyses, successfully recovers these features in LBG spectra. This demonstrates its robustness and adaptability to a broader class of high-redshift background sources.

\subsection{Correlation Function Measurement}\label{subsec:correlations}
We estimate both the \lya auto-correlation of the transmitted‐flux field $\delta_g(\lambda)$ (described in \Cref{subsec:delta_extraction}) and its cross-correlation with the LBG/LAE tracer catalogs introduced in \Cref{sec:dataset}. Pair separations, either between \lya\ pixels in the auto-correlation or between \lya\ pixels and galaxies in the cross-correlation, are binned in the line-of-sight ($\rp$) and transverse ($\rt$) directions. To compute these separations, angular and redshift differences are converted into comoving distances using the fiducial cosmology of~\cite{Planck:2018CosmoParams}. 

For both correlation functions, we adopt $\Delta R = 2\ \hMpc$ bins spanning the range $0 < \rp, \rt < 48\ \hMpc$ for the auto-correlation, and $0 < |\rp|, \rt < 48\ \hMpc$ for the cross-correlations. In the latter case, negative $\rp$ values correspond to configurations where the tracer galaxies are located behind the \lya\ pixels along the line of sight.

The auto‐correlation in a $M\equiv(\rp,\rt)$ bin, is defined as
\begin{equation}
    \xi^{\alpha \times \alpha}_M = \frac{\sum_{i,j\in M} w^\alpha_i\  w^\alpha_j\  \delta_i\  \delta_j}{\sum_{i,j\in M} w^\alpha_i\  w^\alpha_j},
\end{equation}
where $\delta_i \equiv \delta_g(\lambda_i)$, and each \lya pixel weight, $w_i^\alpha$, is
\begin{equation}\label{eq:weights_lya}
    w^\alpha_i \equiv w(\lambda_i) = \left(\frac{1+z_i}{1+2.25}\right)^{\gamma_\alpha-1} \left[ \eta(\lambda_i) \  \tilde{\sigma}^2_{{\rm pipe},g}(\lambda_i) + \eta_{\rm LSS}\  \sigma^2_{\rm LSS}(\lambda_i) \right]^{-1} ,
\end{equation}
where $\tilde{\sigma}^2_{{\rm pipe},g}(\lambda) \equiv \sigma^2_{{\rm pipe},g}(\lambda)/[\overline{F}(\lambda)C_g(\lambda)]^2$, $\gamma_\alpha=2.9$~\citep{SDSS:2004kjl}, $\eta(\lambda)=1$, and $\eta_{\rm LSS}$ modulates the contribution of the intrinsic \lyaf flux variance.\footnote{In \DESIDRIILya, $\eta_{\rm LSS}$ was set to 7.5 based on results from the DESI Early Data Release (EDR) \lya analysis to minimize the quasar \lyaf auto-correlation covariance~\citep{2023RamirezPerez:LyaCatalogEDR,2023JCAP...11..045G}. Here, we fix $\eta_{\rm LSS}=1$ as determining the optimal value for LBG forests requires further study.}

Similarly, the \lya cross‐correlation with the galaxy tracer $X$ is
\begin{equation}
    \xi^{\rm \alpha \times X}_M 
    = \frac{\sum_{i,j \in M} w^\alpha_i\  w^{\rm X}_j\  \delta_i}
           {\sum_{i,j \in M} w^\alpha_i\  w^{\rm X}_j},
\end{equation}
with galaxy tracer weights, $w^{\rm X}_i$, given by
\begin{equation}\label{eq:weights_X}
    w^{\rm X}_i 
    = \left(\frac{1+z_i}{1+2.25}\right)^{\gamma_{\rm X}-1}.
\end{equation}
In this analysis, we adopt $\gamma{\rm X}=1.44$, following the convention commonly used for quasars~\citep{dMdB2019:MgII}.
While more detailed models exist for the redshift and magnitude dependence evolution of dropout-selected galaxies (e.g., Equation 2.7 in~\citep{WilsonWhite2019}) we opt here for this simplified prescription. Incorporating a more refined evolution model is left for future work.

We estimate the covariance matrix of our correlation measurements by dividing the survey footprint into HEALPix pixels with $N_{\rm side}=64$~\citep{Gorski2005}, each covering approximately $0.84\ \deg^2$, which corresponds to a comoving scale of $\sim70\ \hMpc$ at $z=2.7$. We compute the correlation function independently within each HEALPix region and derive the covariance matrix using a bootstrap resampling approach, drawing 10,000 realizations from the set of HEALPix pixel subsamples. The estimated covariance is then smoothed following the same recipe as described in Section 3.3 of \cite{2025JCAP...01..124A}.

Due to the relatively small footprint of our analysis (and hence the limited number of independent samples) the estimated covariance may be underestimated. This limitation could be mitigated in future work by analyzing larger sky areas or by extracting the covariance from several realizations of synthetic LBG spectra datasets.

\subsection{Model of the Correlation Functions}\label{sec:model}
Deriving tight model constraints is not the primary objective of this study. Nevertheless, we include a model of the correlation functions, analogous to that used by \DESIDRIILya, for completeness. Our analysis employs an anisotropic cross-power spectrum model,
\begin{equation}
P^{\rm \alpha\times Y}(\vb{k}, z) = b_\alpha b_{\rm Y} \left(1+\beta_\alpha\mu_k^2\right)\left(1+\beta_{\rm Y}\mu_k^2\right) P_L(k,z) G(\vb{k})F_{\rm NL}(\vb{k}),
\end{equation}
where $Y$ denotes either \lya\ or a galaxy tracer, and $\vb{k} \equiv (k_\parallel,k_\perp)$ is the wave-number vector with magnitude $k=\sqrt{k_\parallel^2+k_\perp^2}$ and $\mu_k=k_\parallel/k$ representing the cosine of the angle between $\vb{k}$ and the line-of-sight. In this model, $P_L(k,z)$ is the linear matter power spectrum computed using \texttt{CAMB}\citep{Lewis:1999CAMB} under the fiducial cosmology of \cite{Planck:2018CosmoParams}, $G(\vb{k}) = \sinc\left(k_\parallel \Delta R/2\right) \sinc\left(k_\perp \Delta R/2\right)$ accounts for the finite binning $\Delta R = 2\ \hMpc$ of the correlation functions, and $F_{\rm NL}(\vb{k})$ captures non-linear corrections at small scales.

In the auto-correlation, $F_{\rm NL}$ accounts for non-linear structure growth, peculiar velocities, and IGM thermal effects (e.g., broadening and pressure) which are incorporated as specified in Equation 3.6 of \cite{Arinyo-i-Prats:2015vqa}, with parameters $q_1=0.8424$, $q_2=0$, $k_v=0.69516\  \hoverMpc$, $a_v=0.4015$, $b_v=1.711$, and $k_p=21.13\  \hoverMpc$ obtained by interpolating the best-fit model from the L160R25 ACCEL2 hydrodynamical simulation at $z\sim2.7$ (see Table A3 of \cite{Chabanier:2024knr}). For the cross-correlation, the combined effects of galaxy velocity dispersion and statistical redshift estimation errors are modeled by a Lorentzian profile:
\begin{equation}\label{eq:lorentzian}
F^{\rm cross}_{\rm NL}(\vb{k}) = \left[1+k^2_\parallel\sigma^2_{z, \rm X} \right]^{-\frac{1}{2}},
\end{equation}
where $\sigma_{z, \rm X}$ is a free parameter representing the velocity dispersion of the tracer due to both peculiar velocities and redshift errors. In addition, to account for systematic redshift offsets that shift the cross-correlation along the line-of-sight, we introduce a free parameter, $\Delta r_{\parallel,\rm X}$, which quantifies this shift.

We model the redshift evolution of the linear biases for \lya\ and galaxies using a power law, $b_{\rm Y}(z) = b_{\rm Y}(z_{\rm eff})\left[(1+z)/(1+z_{\rm eff})\right]^{\gamma_{\rm Y}}$, with $\gamma_\alpha=2.9$ for \lya and $\gamma_{\rm X}=1.44$ for galaxy tracers, as in Sect.~\ref{subsec:correlations}. For galaxy tracers, the RSD parameter is derived from the relation $\beta_{\rm X} = f(z)/b_{\rm X}(z)$, where $f(z) \sim \Omega_m^{0.55}(z)$ is the linear growth rate. In our analysis, $f(z)$ is fixed to its value at $z\sim 2.7$ (i.e, $f\sim 0.98$). Note that this relation is not applied to the \lyaf, for which $\beta_\alpha$ is treated as independent, though we approximate it as constant in redshift for simplicity, despite evidence of redshift dependence~\citep{Arinyo-i-Prats:2015vqa,Chabanier:2024knr}.

Additionally, we account for contributions to the correlation functions from contaminants to the \lyaf such as unmasked High Column Density systems (HCDs), absorption from other atomic transition lines blended within the \lyaf\ (hereafter referred to as metals), and correlated noise. These contributions, if not properly modeled, could bias the \lyaf\ parameter measurements. The procedures used to incorporate these effects are described in \Cref{subsec:contaminants}.

The final power spectrum model is decomposed into multipoles up to $\ell=6$ and transformed into correlation function multipoles via a Hankel transform. The resulting model is then interpolated onto an $(\rp,\rt)$ grid to yield the following forms for the correlation functions:
\begin{align}
\xi^{\alpha\times\alpha}_{\rm Total} &= \xi^{\alpha\times\alpha} + \sum_m \xi^{\alpha\times m} + \xi^{\rm noise}, \\
\xi^{\rm \alpha\times X}_{\rm Total} &= \xi^{\rm \alpha\times X} + \sum_m \xi^{{\rm X}\times m},
\end{align}
where each $m$ labels a distinct metal transition. Finally, distortions in the correlation functions arising from the continuum fitting procedure are incorporated via a distortion matrix, following the approach described in Section III.A of \DESIDRIILya\ and in \citep{Busca2025}.

\subsubsection{Contaminants Model}\label{subsec:contaminants}
\paragraph{High Column Density systems.} HCDs, including Lyman‑limit systems (LLS) and DLAs, are biased tracers of the underlying matter density field.  When blended into the \lyaf, their presence introduces a scale‑dependent excess of large‑scale power into the 3D \lyaf correlations~\citep{McQuinn2011:HCDs,Rogers:2017bmq,Tan2025}, boosting the inferred \lya linear bias and suppressing the RSD parameter~\citep{Font2012:HCDEffect}. To account for these effects, we model the \lya bias and RSD as ``effective” parameters that include a scale‑dependent HCD contribution:
\begin{align}
    b'_\alpha &= b_\alpha + b_{\rm HCD}F_{\rm HCD}(k_\parallel), \\
    b'_\alpha\beta'_\alpha &= b_\alpha\beta_\alpha +b_{\rm HCD}\beta_{\rm HCD}F_{\rm HCD}(k_\parallel),
\end{align}
where $b_{\rm HCD}$ and $\beta_{\rm HCD}$ are the linear bias and RSD parameter associated with the HCD contribution to the transmitted-flux fraction. Note that $b_{\rm HCD}$ is negative and small in this context, as it describes absorption in the transmitted-flux field, rather than the clustering bias of the HCD systems themselves, which is typically $\sim2$~\citep{2012JCAP...11..059F,Perez-Rafols:2017mjf}. The term $F_{\rm HCD}(k_\parallel)=\exp(-L_{\rm HCD} k_\parallel)$ describes the scale-dependent contribution of HCDs to the correlation function, with $L_{\rm HCD}$ representing their typical length scale. 

\paragraph{Other atomic transitions (metals).}
Our pipeline assumes that every absorption in the analysis region arises from \lya but in reality metal absorption blended into the forest introduce spurious correlations most noticeable along the line-of-sight. We include this contamination via metal cross-terms, $\xi^{\alpha\times m}$ in the auto‐correlation and $\xi^{{\rm X}\times m}$ in the cross‐correlation. Given the scales relevant to our analysis, we model only the contribution from the \ion{Si}{III}{1207} transition, treating its linear bias, $b_{\rm \ion{Si}{III}{1207}}$, as a free parameter and fixing its RSD parameter to $\beta_{\rm \ion{Si}{III}{1207}}=0.5$. We neglect metal auto‐correlations and cross-correlation between distinct metals as they are subdominant. Strong transitions such as C\thinspace IV, Si\thinspace IV, and Mg\thinspace II can in principle significantly contribute through their auto-correlation~\citep{KP6s5-Guy}; however, preliminary analyses did not show a robust signal detection.

\paragraph{Correlated noise.} The DESI data reduction pipeline introduces correlated noise between fibers on the same focal‐plane petal~\citep{Spectro.Pipeline.Guy.2023,KP6s5-Guy}, which contaminates the \lyaf auto‐correlation. We model this contribution as an additive term,
\begin{equation}
    \xi^{\rm noise} = a_{\rm noise} \delta^K(\rp)f(\rt),
\end{equation}
where $a_{\rm noise}$ is a free parameter that sets the amplitude of the noise contribution, $\delta^K$ is the Kronecker delta in $r_\parallel$, and $f(r_\perp)$ is a decreasing function of $r_\perp$, normalized such that $f(0)=1$, proportional to the probability that two pixels originate from the same petal.

\section{Results and discussion}\label{sec:results}
\subsection{Correlation functions}\label{subsec:corr_results}
\begin{figure}[!tbp]
    \centering
    \includegraphics[width=\textwidth]{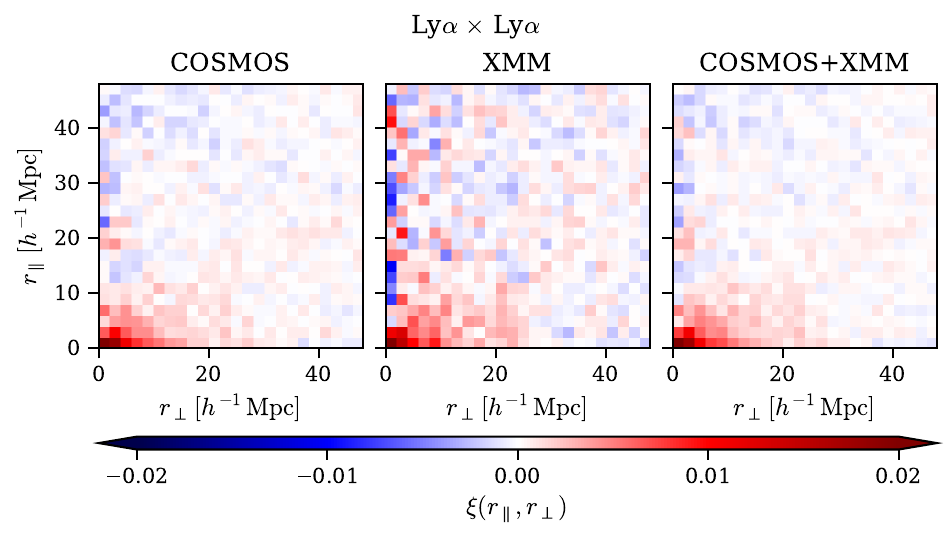}
    \caption{\lyaxlya\ auto‑correlation measured from LBG spectra with $\mathrm{CL}>0.995$ in COSMOS (left), XMM (center), and the combined fields (right), shown over a $(\rp,\rt)$ grid. While some visual differences are apparent between COSMOS and XMM, we have verified that the measurements are statistically consistent; these differences are likely due to noise and statistical fluctuations.}
    \label{fig:auto-correlation_matrix}
\end{figure}

Now we focus on the \lyaf\ auto and cross‑correlations with LBG and LAE positions results. \Cref{fig:auto-correlation_matrix} shows the LBG forest \lyaxlya\ auto‑correlation as a function of $\rp$ and $\rt$ in the COSMOS (left) and XMM (center) fields, using LBG spectra with CNN confidence level $\mathrm{CL}>0.995$. The effective redshift for this measurement is $z_{\rm eff}=2.70$. Some visual discrepancies appear between the COSMOS and XMM measurements. However, these are likely driven by statistical fluctuations and noise. We have verified that the measurements are consistent within uncertainties. Because the mean rest-frame continuum (see \Cref{subsec:delta_extraction}), signal-to-noise ratios, magnitude ranges, and redshift distributions are consistent across both fields, we combine them to improve statistical power. The right panel shows results for the combined sample.

\begin{figure}
    \centering
    \includegraphics[width=\textwidth]{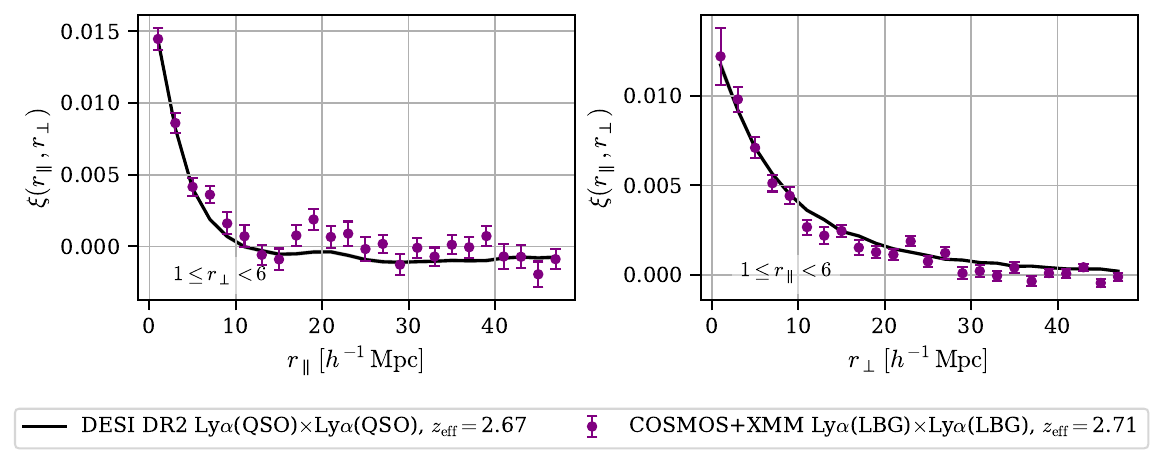}
    \caption{\lyaxlya auto-correlation from the COSMOS+XMM LBG forest combined sample as a function of $r_\parallel$ (left) and $r_\perp$ (right), averaged over the $1-6\ \hMpc$ range of the complementary coordinate. The black line shows the results from the \DESIDRIILya\ quasar forest sample where forest pairs have been restricted to match the effective redshift of our LBG forest signal. The (almost indistinguishable) shaded black region corresponds to the 1$\sigma$ uncertainty. }
    \label{fig:auto-bins}
\end{figure}

For illustrative purposes, \Cref{fig:auto-bins} displays the same \lya auto-correlation signal from COSMOS+XMM, averaged over the first five bins in the transverse (left) and parallel (right) directions. For comparison, we include the \lya auto-correlation measurement from $\sim$766,000 quasar forests in the \DESIDRIILya\ sample, where we restricted forest pairs to match the effective redshift of our LBG forest signal. Throughout this section, we adjust our notation to Ly$\alpha$(LBG) and Ly$\alpha$(QSO) to differentiate the origin of the \lyaf signal.

Our Ly$\alpha$(LBG)$\times$Ly$\alpha$(LBG) auto-correlation measurement represents the first detection of the \lya forest auto-correlation signal using exclusively LBG spectra, at an effective redshift of $z_{\rm eff} \sim 2.7$. While previous works have included LBG forests~\cite[e.g.,][]{Momose2021,Momose2021LAE,Newman2024,Zhang2025}, they primarily combined them with quasar forests to study Ly$\alpha$–galaxy cross-correlations. 

Our measurement is consistent within uncertainties with the Ly$\alpha$(QSO)$\times$Ly$\alpha$(QSO) correlation obtained from the DESI DR2 quasar sample when forest pairs are restricted to match the same effective redshift ($z_{\rm eff}$). We do not observe any additional systematic effects beyond those already present in the quasar analysis. Furthermore, we verified that excluding DLA masking in the DESI DR2 quasar forests introduces only minor differences, well within the uncertainties of our LBG forest measurements.

We have also confirmed that the distortion introduced by the continuum-fitting procedure is comparable between the quasar and LBG-based measurements at the scales relevant to our study. This consistency supports the expectation that both signals should closely match. A notable exception is observed in the \ion{Si}{III}{1207} feature at $\rp \sim 20\ \Mpc$ at transverse distances close to the line-of-sight (see \Cref{subsec:contaminants}), which appears more prominent in the LBG sample. Interestingly, the strength of this feature diminishes as we increase the CNN confidence threshold (see Appendix~\ref{appendix:posteriors}), suggesting that it likely arises from contamination by objects misclassified as LBGs, redshift errors outliers, or from spectral noise.

\begin{figure}[!tbp]
    \centering
    \includegraphics[width=0.79\textwidth]{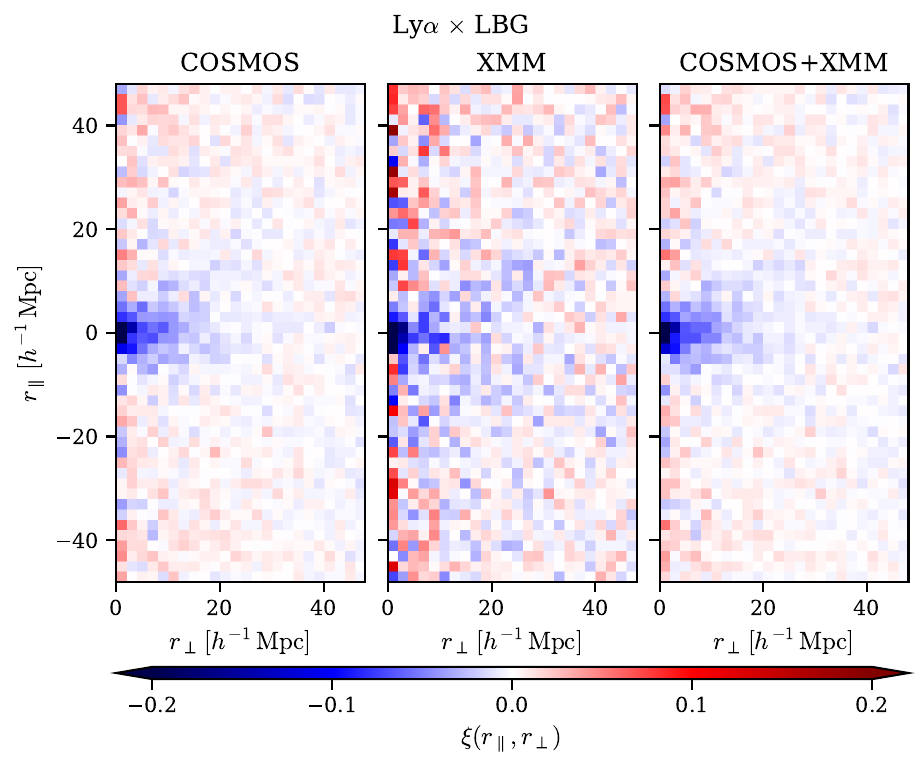}
    \includegraphics[width=0.79\textwidth]{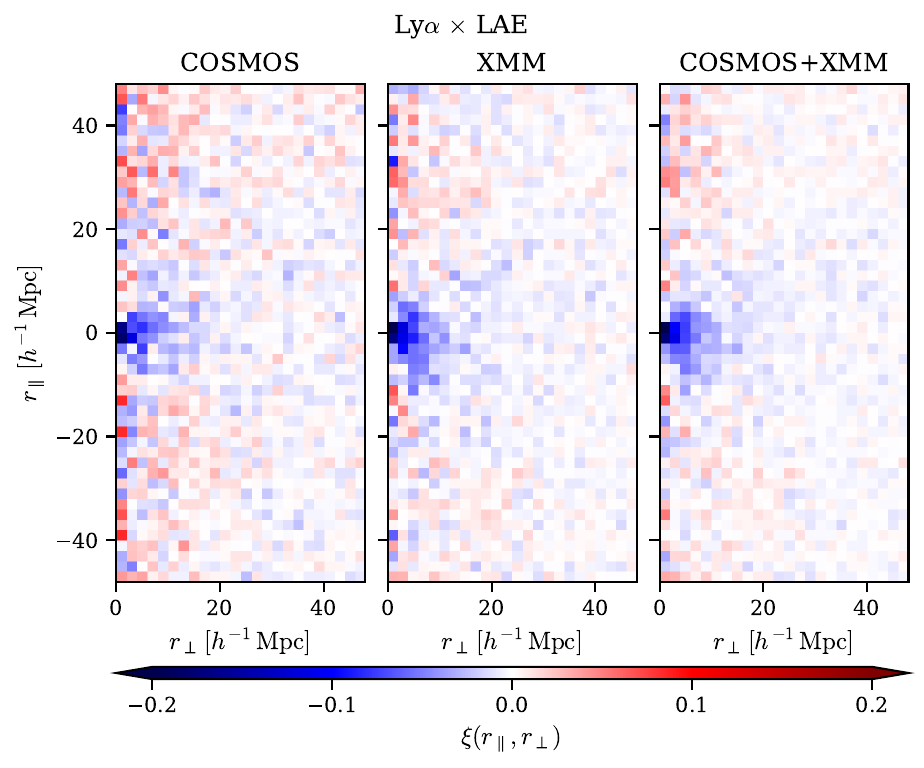}
    \caption{\lyaxlbg\ (top) and \lyaxlae\ (bottom) cross-correlations in COSMOS (left), XMM (center), and the combined (right) samples, shown over a $(\rp,\rt)$ grid.}
    \label{fig:cross_matrix}
\end{figure}

\begin{figure}
    \centering
    \includegraphics[width=\textwidth]{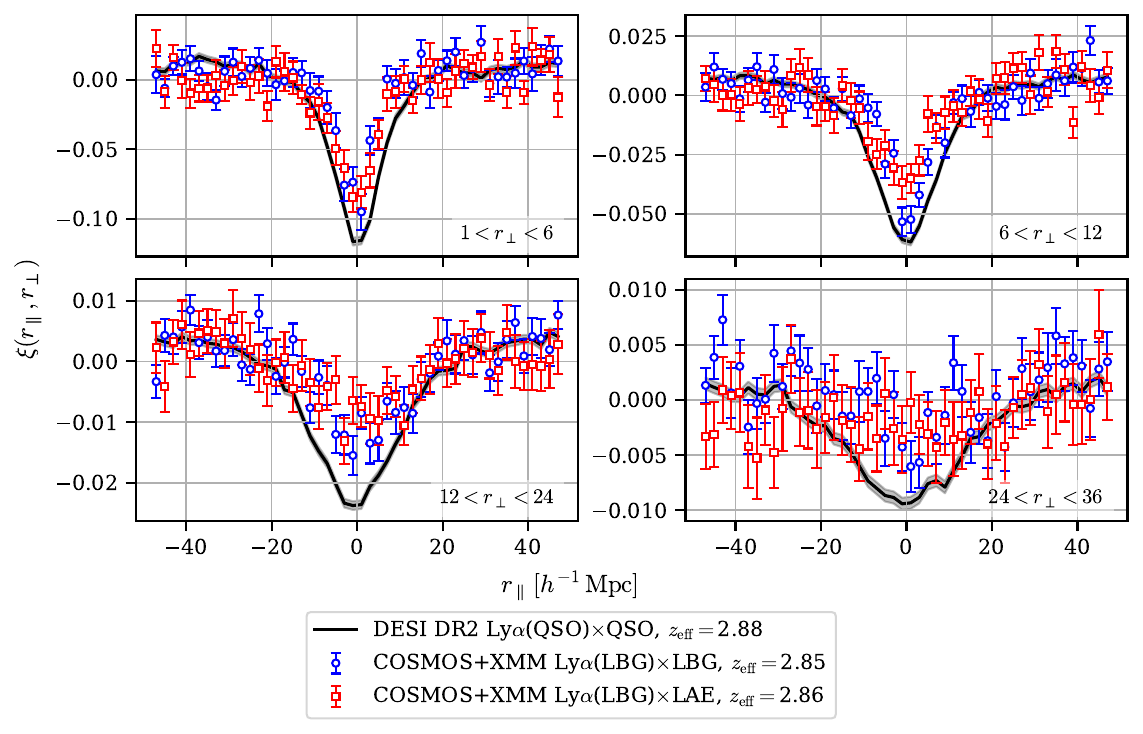}
    \caption{Measured cross-correlation between the \lyaf in LBG spectra and the positions of LBGs (blue unfilled dots) and LAEs (red unfilled squares) from the COSMOS+XMM combined sample as a function of $\rp$, averaged over four distinct $\rt$ bin slices. The black line shows the cross-correlation between the \lyaf from quasars and quasar tracer positions from the \DESIDRIILya\ sample with the shaded region corresponding to 1$\sigma$ uncertainties. The DESI DR2 forest–quasar pairs have been restricted to ensure that the Ly$\alpha$(QSO)$\times$QSO signal has an effective redshift comparable to that of our Ly$\alpha$(LBG)$\times$LBG and Ly$\alpha$(LBG)$\times$LAE signals.}
    \label{fig:rtbins_cross}
\end{figure}

Similarly, \Cref{fig:cross_matrix} presents the Ly$\alpha$(LBG)$\times$LBG (top) and Ly$\alpha$(LBG)$\times$LAE (bottom) cross‑correlations on the $(r_\parallel,r_\perp)$ grid for COSMOS, XMM, and the combined dataset. \Cref{fig:rtbins_cross} shows the same cross‑correlation signals averaged over four $\rt$ slices. We include the Ly$\alpha$(QSO)$\times$QSO cross-correlation results from the aforementioned DESI DR2 quasar forests and the position of quasar tracers, where forest–quasar pairs were restricted to match the effective redshift of our LBG and LAE cross-correlation signals.

We observe that the Ly$\alpha$(LBG)$\times$LBG and Ly$\alpha$(LBG)$\times$LAE cross‑correlation functions differ from the Ly$\alpha$(QSO)$\times$QSO signal, showing lower amplitudes at small-scales. This reflects the fact that LBGs~\citep{2013VLT:LBG,Jose:2013LBGs} and LAEs~\citep{2016VLT:LAE,White:2024LAE,Ebina2025} reside in lower-mass halos, resulting in reduced clustering amplitude (lower linear bias) and smaller intrinsic velocity dispersions compared to quasars~\citep{Eftekharzadeh:2015ywa,Chehade:2016xoh,deBeer:2023ccl,Eltvedt:2024zas}. Moreover, redshift estimate errors introduce additional broadening of the profile. Both the Ly$\alpha$(LBG)$\times$LBG and Ly$\alpha$(LBG)$\times$LAE cross‑correlations peak at $r_\parallel \sim 0\ \hMpc$, especially in the $1 < \rt < 6\ \hMpc$ and $6 < \rt < 12\ \hMpc$ slices, confirming that our systematic redshift offset corrections (see Appendix~\ref{appendix:redshift_correction}) are appropriate.

Our detection of the LBG forest auto-correlation, together with its cross-correlations with LBG and LAE positions, shows that correlations based on LBG forests alone can be robustly measured. In contrast to the relatively high SNR LBG spectra used in LATIS and CLAMATO, our measurement relies on lower SNR ($\sim0.35$ \AA$^{-1}$) data, providing the first direct demonstration of the viability of LBG forests in the context of wide-area surveys such as DESI. These results indicate that LBG forests can complement quasar-based studies by extending clustering measurements to higher source densities and redshifts. Looking ahead, larger datasets such as those anticipated from DESI-II~\citep{DESI-II:2022} will be required to test this approach at higher precision and to explore whether the BAO feature can be detected from LBG forests. We discuss these prospects further in \Cref{sec:prospects}.

\subsection{Model parameter constraints}\label{subsec:param_contraints}
As noted in \Cref{sec:model}, deriving tight model constraints is beyond the primary scope of this work. This is primarily because the statistical power of our measurements is insufficient, and also because we adopt a simplified model that does not fully capture the underlying physics on the small scales considered here, potentially biasing the results. Nevertheless, we perform model fitting for completeness and emphasize that \emph{the following results should be interpreted with caution}. 

For the remainder of this section, we focus only on the \lya\ and LBG-related parameters. A more complete discussion of the other model parameters and their posterior distributions can be found in Appendix~\ref{appendix:posteriors}.

We adopt a Gaussian likelihood and use the \texttt{Vega} package~\github{https://github.com/andreicuceu/vega} to model the correlation functions. Best-fit parameter values and statistics are estimated using the \texttt{iminuit} minimizer~\citep{James:1975dr,iminuit}, while posterior distributions are sampled via the \texttt{Polychord} nested sampler~\citep{Handley:2015fda,Handley:2015Polychord}, both integrated within the \texttt{Vega} interface. We adopt wide flat priors for all free parameters, except for the HCD redshift-space distortion parameter, for which we impose a Gaussian prior of $\beta_{\rm HCD} = 0.5 \pm 0.2$. This is broader than the prior used in \DESIDRIILya\ to allow for possible redshift evolution of $\beta_{\rm HCD}$, given that our analysis targets higher redshifts.

We carry out auto+cross-correlation joint fits, also accounting for their cross-covariance. We do not present auto or cross‐only fits, as their statistical power is significantly lower than that of the combined fit. We also attempted additional fits based on the \lyaxlae cross-correlation; however, these proved unstable because most of our LAE sample lies in the XMM field, whereas the bulk of our \lyaf\ statistics come from COSMOS (see \Cref{tab:statistics}); therefore, we omit these results from this work. 

\begin{table}[t]
\centering
\caption{Parameter constraints at 68\% confidence from joint auto+cross‐correlation fits, applied to the combined COSMOS+XMM LBG sample. Our full model is described in \Cref{sec:model}. We report the best-fit model $\chi^2_{\rm min}$ value divided by the degrees of freedom (D.o.F). Results are shown for several fitting ranges in comoving separation.  Entries given as $\mu\pm\sigma$ denote Gaussian posterior distributions, while those of the form $\mu^{+\sigma_1}_{-\sigma_2}$ indicate non‐Gaussian posteriors.}\label{tab:fit_results_combined}
\resizebox{\textwidth}{!}{%
\begin{tabular}{lccccccc}

Fitting range & $b_{\alpha}$ & $\beta_{\alpha}$ & $b_{\rm LBG}$ & $\sigma_{z,\rm LBG}\  [\hMpc]$& $\Delta r_{\parallel,\rm LBG}\  [\hMpc]$ & $\chi^2_{\rm min}/{\rm D.o.F}$ \\
\toprule

\multicolumn{8}{l}{\combinedLBG \rule{0pt}{4ex}} \\
$1<r<28$ & $-0.152^{+0.017}_{-0.023}$ & $1.46^{+0.23}_{-0.36}$ & $1.90\pm0.13$ & $<0.780$ & $0.04\pm0.18$ & $461.29/(462 - 10) = 1.02$ \\
$6<r<28$ & $-0.187^{+0.028}_{-0.042}$ & $1.10^{+0.16}_{-0.45}$ & $1.85\pm0.16$ & $<0.790$ & $-0.37\pm0.36$ & $425.71/(438 - 10) = 0.99$ \\
$10<r<28$ & $-0.137\pm0.043$ & $1.99^{+0.42}_{-1.1}$ & $1.93\pm0.24$ & $<1.350$ & $-1.09\pm0.54$ & $382.31/(402 - 10) = 0.98$ \\
$1<r<34$ & $-0.167^{+0.015}_{-0.020}$ & $1.24^{+0.18}_{-0.26}$ & $1.88\pm0.13$ & $<0.760$ & $0.04\pm0.19$ & $711.19/(684 - 10) = 1.06$ \\
$6<r<34$ & $-0.209^{+0.019}_{-0.031}$ & $0.85^{+0.13}_{-0.26}$ & $1.80\pm0.16$ & $<0.800$ & $-0.29\pm0.36$ & $672.01/(660 - 10) = 1.03$ \\
$10<r<34$ & $-0.164^{+0.034}_{-0.046}$ & $1.39^{+0.21}_{-0.68}$ & $1.90\pm0.25$ & $<1.220$ & $-0.93\pm0.54$ & $629.60/(624 - 10) = 1.03$ \\
\end{tabular}
}
\end{table}

 \Cref{tab:fit_results_combined} presents the parameter constraints derived from our combined COSMOS+XMM LBG sample auto+cross-correlation fits when varying the comoving separation fitting range. From these results we draw the following conclusions:
\begin{itemize}[noitemsep]
  \item \textbf{\lya bias parameters ($b_\alpha$, $\beta_\alpha$):}  These parameters exhibit significant variation depending on the chosen fitting range, with $b_\alpha$ spanning from $-0.209$ to $-0.137$ and $\beta_\alpha$ ranging from 0.85 to 1.99. While including smaller-scale separations tightens statistical uncertainties, it also increases susceptibility to non-linear effects. On the other hand, within the limits of our dataset, including larger-scale separations does not improve parameter stability. The degeneracy of the \lya parameters with the HCD linear bias parameter, in addition to the limited statistical power of our current dataset further complicates the robust disentanglement of \lya\ and HCD contributions. Taken together, these factors prevent a reliable measurement of $b_\alpha$ and $\beta_\alpha$ within the scope of this analysis.
  \item \textbf{LBG bias ($b_{\rm LBG}$):} This parameter remains stable across all fitting ranges explored ranging from 1.80 to 1.93. We note that the inferred values are systematically lower than those reported in previous LBG clustering analyses at comparable redshifts~\cite[e.g.,][]{2013VLT:LBG,Ruhlmann-Kleider:2024LBG}, which typically find $b_{\rm LBG} \sim 2.6$–$3.5$. We attribute this discrepancy primarily to sample purity. As will be discussed later in this section, the inferred $b_{\rm LBG}$ increases as a function of the CNN confidence level threshold used for selecting LBGs, indicating a strong dependence on contamination, classification reliability, and redshift errors outliers.
  \item \textbf{Velocity dispersion ($\sigma_{z,\rm LBG}$):} We are unable to robustly determine a central value for this parameter, obtaining only upper limits that vary with the fitting range but consistently remain below $\sim 2.7\ \hMpc$ at the 95\% confidence level. We also note that this parameter is stable against CNN confidence level. This upper bound is relatively smaller than the $\sim 3.7\ \hMpc$ velocity dispersion measured for \DESIDRIILya\ quasars~\citep{DESI.DR2.BAO.lya}.
  \item \textbf{Redshift offset ($\Delta r_{\parallel,\rm LBG}$):} Across all fitting ranges, the inferred redshift offset is consistent with zero, with the largest deviation reaching only $\sim2\sigma$ when small scales ($r < 10\ \hMpc$) are excluded. This supports the effectiveness of the redshift estimate correction described in Appendix~\ref{appendix:redshift_correction}.
\end{itemize}

\begin{table}[!tbp]
\centering
\caption{Parameter constraints at 68\% confidence from \combinedLBG joint fits from alternative analysis and modeling strategies when fitting over the $1< r<28\ \hMpc$ range. We adopt as a reference the results from the corresponding fitting range from \Cref{tab:fit_results_combined}. All our variations have an effective redshift $z_{\rm eff}= 2.71$, except for the $\mathrm{CL}>0.999$ variation which has an effective redshift $z_{\rm eff} = 2.72$.}\label{tab:fit_variations}

\resizebox{\textwidth}{!}{%
\begin{tabular}{lcccccccc}
  Variation & $b_{\alpha}$ & $\beta_{\alpha}$ & $b_{\rm LBG}$ & $\sigma_{z,\rm LBG}\  [\hMpc]$& $\Delta r_{\parallel,\rm LBG}\  [\hMpc]$ & $\chi^2_{\rm min}/{\rm D.o.F}$ \\
\toprule

\textbf{Reference \rule{0pt}{4ex}} & $-0.152^{+0.017}_{-0.023}$ & $1.46^{+0.23}_{-0.36}$ & $1.90\pm0.13$ & $<0.780$ & $0.04\pm0.18$ & $461.29/(462 - 10) = 1.02$ \\
\multicolumn{9}{l}{\textbf{CNN Confidence level} \rule{0pt}{4ex}} \\
$\mathrm{CL}>0.97$ & $-0.156^{+0.018}_{-0.022}$ & $1.39^{+0.21}_{-0.36}$ & $1.322^{+0.092}_{-0.10}$ & $<0.760$ & $-0.09\pm0.17$ & $504.20/(462 - 10) = 1.12$ \\
$\mathrm{CL}>0.99$ & $-0.186^{+0.016}_{-0.021}$ & $0.97^{+0.16}_{-0.23}$ & $1.71\pm0.12$ & $<0.760$ & $-0.07\pm0.18$ & $492.76/(462 - 10) = 1.09$ \\

$\mathrm{CL}>0.999$ & $-0.168^{+0.019}_{-0.032}$ & $1.36^{+0.20}_{-0.41}$ & $2.18\pm0.15$ & $<0.750$ & $-0.29\pm0.19$ & $514.74/(462 - 10) = 1.14$ \\

\multicolumn{9}{l}{\textbf{Bin width $\Delta R$} \rule{0pt}{4ex}} \\
$\Delta R = 1\ \hMpc$ & $-0.163^{+0.017}_{-0.022}$ & $1.29^{+0.20}_{-0.31}$ & $1.79\pm0.12$ & $<0.960$ & $-0.16\pm0.16$ & $2095.91/(1854 - 10) = 1.14$ \\
$\Delta R = 3\ \hMpc$ & $-0.174^{+0.020}_{-0.026}$ & $1.25^{+0.20}_{-0.35}$ & $1.86\pm0.13$ & $<0.650$ & $-0.02\pm0.19$ & $243.75/(207 - 10) = 1.24$ \\
$\Delta R = 4\ \hMpc$ & $-0.186^{+0.019}_{-0.024}$ & $1.09^{+0.20}_{-0.29}$ & $1.74\pm0.15$ & $<1.060$ & $-0.03\pm0.26$ & $133.96/(117 - 10) = 1.25$ \\

\multicolumn{9}{l}{\textbf{Alternative model or priors} \rule{0pt}{4ex}} \\
Gaussian $F_{\rm NL}^{\rm cross}$ & $-0.154^{+0.017}_{-0.021}$ & $1.46^{+0.22}_{-0.37}$ & $1.89\pm0.13$ & $<1.130$ & $0.04\pm0.20$ & $461.29/(462 - 10) = 1.02$ \\
Gaussian prior $\beta_\alpha$ & $-0.148^{+0.007}_{-0.01}$ & $1.540\pm0.095$ & $1.89\pm0.13$ & $<0.790$ & $0.04\pm0.19$ & $461.80/(462 - 10) = 1.02$ \\
$F_{\rm NL}^{\rm auto}$ parameters & $-0.154^{+0.017}_{-0.021}$ & $1.41^{+0.22}_{-0.34}$ & $1.90^{+0.12}_{-0.14}$ & $<0.800$ & $0.04\pm0.19$ & $461.99/(462 - 10) = 1.02$ \\
Fixed $\beta_\alpha$ and HCDs & $-0.130\pm0.004$ & $1.71$ & $1.89\pm0.13$ & $<0.740$ & $0.04\pm0.20$ & $463.92/(462 - 6) = 1.02$ 
\end{tabular}}
\end{table}

We further assess the robustness of our measurements by exploring a range of alternative analysis and modeling choices. In \Cref{tab:fit_variations}, we summarize the resulting parameter constraints, taking as a baseline the \combinedLBG joint fit over the range $1< r<28\ \hMpc$. The variations explored include:
\begin{itemize}[noitemsep]
  \item \textbf{CNN confidence level:} We explore alternative LBG selection thresholds of $\mathrm{CL}>0.97$, 0.99, and 0.999, relative to our baseline of $\mathrm{CL}>0.995$. Increasing the threshold systematically raises the inferred LBG bias, $b_{\rm LBG}$, consistent with higher sample purity, $p$, and a reduced fraction of catastrophic redshift errors, $f$. \RuhlmannLBG\ report $p \sim 0.94$ and $f \sim 0.10$ at $\mathrm{CL}>0.995$. Independently, our comparison with the COSMOS spectroscopic compilation in \Cref{subsec:redshift_measurements} indicates a similar redshift outlier fraction. Since our sample extends to fainter magnitudes than either the \RuhlmannLBG or the cross-matched COSMOS samples, we expect slightly lower purity and a correspondingly higher fraction of catastrophic redshift errors, which would further suppress $b_{\rm LBG}$. In contrast, we observe no clear trend in the \lya\ linear bias $b_\alpha$ with $\mathrm{CL}$, partly due to its degeneracy with $\beta_\alpha$ and the HCD parameters. This degeneracy is discussed in Appendix~\ref{appendix:posteriors}, where we also present a complementary test that confirms a similar dependence of $b_\alpha$ on the $\mathrm{CL}$ threshold. Together, these results highlight the need for a more refined treatment of interlopers and catastrophic redshift errors in future large-scale clustering analyses with LBG forest.
  \item \textbf{Bin width $\Delta R$:} We vary the bin width $\Delta R$ between $1$ and $4\ \hMpc$, compared to our baseline of $2\ \hMpc$. Focusing on the redshift estimate offset parameter $\Delta r_{\parallel,\rm LBG}$, we find values consistent with zero across all choices of $\Delta R$, confirming the robustness of our previous results. In contrast, the parameters $b_\alpha$, $\beta_\alpha$, $b_{\rm LBG}$, and $\sigma_{z,{\rm LBG}}$ show noticeable sensitivity to the bin width.
  \item \textbf{Alternative parametrization for the LBG redshift dispersion:} For the effect of velocity dispersion and redshift errors on the cross-correlation, we adopt a Gaussian model, $F^{\rm cross}_{\rm NL}(\vb{k}) = \exp(-\frac{1}{2} k^2_\parallel \sigma^2_{z,\rm X})^{-\frac{1}{2}}$, instead of the Lorentzian profile introduced in \Cref{eq:lorentzian}. The inferred value in this case is $\sigma^2_{z,\rm LBG} < 1.71\ \hMpc$ at 95\% the confidence level, consistent with our previous findings.
  \item \textbf{Alternative constraints on $\beta_\alpha$ and $F^{\rm auto}_{\rm NL}$ parameters:} In a first fitting variation, we impose a Gaussian prior $\beta_\alpha=1.56\pm0.1$ based on results from the ACCEL2 hydrodynamic simulations~\citep{Chabanier:2024knr}. In a second variation, we use an alternative set of (fixed) parameters for the $F^{\rm auto}_{\rm NL}$ auto-correlation small-scales correction.\footnote{Specifically: $q_1=0.7768$, $k_v=1.26735\, \hoverMpc$, $a_v=0.6094$, $b_v=1.645$, and $k_p=20.09\, \hoverMpc$ obtained by interpolating the results from Table 7 of \cite{Arinyo-i-Prats:2015vqa} at $z_{\rm eff}\sim 2.7$} Both variations again highlight the important sensitivity of our best-fit values for $b_{\alpha}$ as a function of the fitting strategy details, while the inferred values for $b_{\rm LBG}$ and redshift error parameters are stable.
\item \textbf{Fixed $\beta_\alpha$ and HCD parameters:} We fix the \lya\ RSD parameter to $\beta_\alpha = 1.71$ and the HCD parameters to $b_{\rm HCD} = -0.092$, $\beta_{\rm HCD} = 0.56$, and $L_{\rm HCD} = 5.39\ \hMpc$. These values were derived from fits to the \lya\ correlation functions of \DESIDRIILya\ quasars, where forest–forest and forest–quasar pairs were restricted to match the effective redshift of our LBG sample, and DLAs were not masked, in order to replicate the conditions of the LBG forest sample. We allow $b_\alpha$ to vary, as it is expected to be affected by sample purity and the redshift outlier fraction, and therefore may deviate from the quasar-based result. Taking the quasar-based measurement of $b_\alpha = -0.144 \pm 0.016$ as a reference, the value inferred from the LBG \lya\ forest in this variation is $\sim$10\% smaller in amplitude. This is consistent with the expected redshift outlier fraction in our dataset and provides an independent cross-check of the trends identified in the CNN confidence-level variations. While the effect of catastrophic redshift errors appear to cause the majority of this difference, contributions from statistical fluctuations and the higher noise of LBG spectra compared to quasars cannot be ruled out.
\end{itemize}

\section{Prospects for Cosmology}\label{sec:prospects}
Our detection of the \lyaf auto and cross-correlations using a relatively low-SNR and small sample of LBG spectra from DESI observations demonstrates the feasibility of performing large-scale clustering analyses with the \lyaf from dense LBG samples in future wide-area spectroscopic surveys. To quantify the cosmological potential of such measurements, we present forecasts for the expected precision on Baryon Acoustic Oscillation (BAO) and Full-Shape (FS) parameters of the correlation functions. We consider a future survey covering 5,000~$\deg^2$ with a surface density of 1,000 spectroscopically confirmed LBGs per~$\deg^2$, assuming target selection similar to that of our COSMOS sample. The expected performance is evaluated using two complementary approaches: a forecast based on synthetic LBG spectra (\Cref{subsec:mock_forecast}) and a Fisher-matrix analysis (\Cref{subsec:fisher_forecast}). For context, we also include analogous quasar-based forecasts over a 5,000~$\deg^2$ footprint of the completed DESI \lya\ quasar survey, allowing a direct comparison with the LBG-based results.

We define the parallel and transverse BAO dilation parameters as
\begin{align}
\ap &= \frac{D_H(z_{\rm eff})/r_d}{\bigl[D_H(z_{\rm eff})/r_d\bigr]_{\rm fid}}, \\
\at &= \frac{D_M(z_{\rm eff})/r_d}{\bigl[D_M(z_{\rm eff})/r_d\bigr]_{\rm fid}},
\end{align}
where $D_H$ is the Hubble distance, $D_M$ the comoving angular diameter distance, and $r_d$ the sound horizon at drag epoch. The subscript “fid” denotes the values given by the Planck 2018 fiducial cosmology~\citep{Planck:2018CosmoParams}. Alternatively, we define the isotropic dilation parameter, $\aiso$~\cite{2025JCAP...01..124A} and the Alcock-Paczynski distortion parameter, $\aAP$, as
\begin{align}
\aiso &= \frac{D^{0.55}_H(z_{\rm eff})D^{0.45}_M(z_{\rm eff})/r_d^2}{\bigl[D^{0.55}_H(z_{\rm eff})D^{0.45}_M(z_{\rm eff})/r_d^2\bigr]_{\rm fid}} = \ap^{0.55}\at^{0.45}\, \\
\aAP &= \frac{D_M(z_{\rm eff})/D_H(z_{\rm eff})}{\bigl[D_M(z_{\rm eff})/D_H(z_{\rm eff})\bigr]_{\rm fid}} = \frac{\at}{\ap},
\end{align}
where the exponent weights in $\aiso$ correspond to the optimal combination found in the \DESIDRIILya\ analysis. While these coefficients are appropriate for quasar \lyaf analyses, the optimal values for LBG \lyaf may differ. This will be clarified once BAO measurements from the LBG forest become available.

As in \Cref{sec:results}, we adopt the notation Ly$\alpha$(LBG) and Ly$\alpha$(QSO) throughout this section to explicitly differentiate between the \lyaf signal extracted from LBG spectra and that from quasars.

\subsection{Forecast with LBG Synthetic Spectra}\label{subsec:mock_forecast}
For the LBG forecasts, we construct a single synthetic dataset of 1,000 LBG/$\deg^2$ over a 5,000 deg$^2$ footprint, employing the following methodology.
\begin{enumerate}[noitemsep]
\item \textbf{Density field simulations:} We use the Quasi-Linear implementation of \texttt{CoLoRe}~\github{https://github.com/damonge/CoLoRe}~\citep{Ramirez2022,Y3.lya-s1.Casas.2025} to generate a Gaussian random field matching an input matter power spectrum, and draw LBG positions via Poisson sampling assuming a density distribution, $n(z)$, similar to our COSMOS sample, rescaled to 1000 LBGs/$\deg^2$ (see \Cref{fig:dn_dz_lbg_qso}). For simplicity, we assume a quasar-like linear bias evolution model.

At this same stage, we generate density and velocity fields along the line-of-sight (also referred to as skewers) associated with each mock LBG. At last, we use \texttt{LyaCoLoRe}~\github{https://github.com/igmhub/LyaCoLoRe}~\citep{Farr2020_LyaCoLoRe} to post-process these skewers into a transmitted flux fraction field via a Fluctuating Gunn-Petterson Approximation (FGPA), also incorporating small-scale corrections and redshift-space distortions.
\item \textbf{Synthetic Spectra generation:} Each mock LBG is assigned a random $r$-band magnitude drawn from the observed COSMOS distribution (in redshift and magnitude). Then, we use the \texttt{quickquasars} script~\github{https://github.com/desihub/desisim/blob/main/py/desisim/scripts/quickquasars.py}~\citep{2024Herrera:Quickquasars} to generate mock spectra, using a common LBG continuum template (redshifted and magnitude-scaled, accordingly). We use \texttt{specsim} package~\github{https://github.com/desihub/specsim}~\citep{Kirkby:2016} to introduce DESI-like instrumental noise assuming two‐hour nominal dark‐time exposures (no moon, 1.1" seeing, airmass 1.0) for each LBG. This configuration results in an average in forest SNR of 0.33 \AA$^{-1}$ for our mock spectra. We do not include contamination due to metals, HCDs, or redshift errors. 
\end{enumerate}

For the quasar forecasts, we generate a corresponding synthetic spectra dataset (also without contaminants) that emulates the characteristics expected for the completed DESI \lya\ survey. The procedure follows the same methodology used for the mocks in previous DESI \lyaf\ analyses~\cite[e.g.,][]{2024Herrera:Quickquasars, Y3.lya-s1.Casas.2025, KP6s6-Cuceu}.

Then, we measure the auto and cross-correlations and associated covariance matrices, following the same recipe described in \Cref{sec:analysis}, adopting a $\Delta R = 4\ \hMpc$ grid up to $0<\rp,\rt<200\ \hMpc$ separations for the auto-correlations and $0<\abs{\rp},\rt<200\ \hMpc$ for the cross-correlations. We compute the covariance by sub-sampling into \path{nside=16} HEALpix pixels and account for cross-covariance between the auto and cross-correlations. Due to computational limitations arising from the high LBG target density, we restrict the LBG forest analysis to a 1,000~$\deg^2$ section and rescale the covariance by a factor of 5 to approximate that of the full 5,000~deg$^2$ footprint. For the quasar analysis, we rescale the covariance matrix by a factor of $\sqrt{14,000/5,000}$ to account for the difference in footprint area between the completed DESI survey and the considered 5,000~$\deg^2$ portion.

At last, we use the \emph{forecast mode} of the \texttt{Vega} code to extract the expected BAO uncertainties. This mode generates a noiseless synthetic data vector representing the correlation function predicted by a given fiducial model. Forecasted errors are then obtained from the covariance matrix via the Hessian of the likelihood function, ensuring that all survey configurations are evaluated under consistent assumptions. We adopt the best-fit model from the \DESIDRIILya\ analysis as our fiducial template; for the LBG forecasts, we exclude quasar-specific terms and include redshift evolution of the \lya\ parameters around $z \sim 2.7$. The BAO scale parameters are fixed to $\ap = \at = 1$. The mock-based forecast results are reported in \Cref{tab:forecasts}: the first three rows correspond to LBG-based correlations, and the next three (below the dashed line) to their quasar-based analogues.

As a complementary forecast, we use the same mock covariance matrices to estimate Alcock–Paczynski constraints from a Ly$\alpha$ Full Shape (FS) analysis, under the same survey setup and fiducial model. We follow the analysis methodology presented in \cite{Cuceu:2021hlk}, previously applied to both mock~\citep{Cuceu:2022brl} and observational data~\citep{Cuceu:2022wbd,Cuceu:LYAFS}. This approach extends the information extracted beyond the BAO peak by including a wider range of scales, leading to significantly improved constraints on the AP parameter.

For the LBG case, we perform FS forecasts using three different minimum separation thresholds for the fit, ranging from an optimistic case ($r > 10\ \hMpc$) to a conservative case ($r > 50\ \hMpc$), while for quasars we only consider the $r > 30\ \hMpc$ case. The results for the intermediate, and most realistic, choice of $r > 30\ \hMpc$ are shown in the final column of \Cref{tab:forecasts}. For the optimistic and conservative configurations on LBGs, we obtain combined auto+cross-correlation constraints of $\sigma^{\textrm{FS}}_{\aAP} = 0.46\%$ and $0.84\%$, respectively.

\subsection{Fisher Forecast}\label{subsec:fisher_forecast}
To complement our mock-based BAO forecast (\Cref{subsec:mock_forecast}), we carry out a Fisher matrix forecast under identical survey conditions. This analysis follows the formalism developed by \cite{McDonaldForecast} and subsequently applied in \cite{DESI2016a.Science,FontForecast}. The forecast is implemented using the publicly available \texttt{lyaforecast} code~\github{https://github.com/igmhub/lyaforecast}. This approach allows for fast and consistent evaluation of the expected BAO precision across different tracer combinations and survey configurations. Unlike the mock-based forecasts, we do not include Full-Shape constraints here, as this functionality is not currently available in \texttt{lyaforecast}.

The Fisher matrix $F_{ij}$ is related to the covariance of our dataset by the relation \citep{Tegmark97}:
\begin{equation}
    F_{ij} = \frac{1}{2}\mathrm{Tr}\left(C^{-1}\frac{\delta C}{\delta \theta_i}C^{-1}\frac{\delta C}{\delta \theta_{j}}\right), 
\end{equation}
where $\theta$ are a set of parameters, and $C(\theta)=\langle\delta_i\delta_j\rangle$, for a Gaussian distributed data vector $\delta$ with $\langle\delta\rangle=0$. Then, the best possible unbiased estimates on $\theta$, given by the Cramer-Rao lower bound \citep{RAOinformation}, are related to the Fisher matrix via:
\begin{equation}\label{eq:cramer_rao}
    \sigma_i \geq (F^{-1})_{ii} , 
\end{equation}
where $\sigma_i$ is the variance on the estimate of parameter $\theta_i$. 

Following \cite{McDonaldForecast}, we construct our covariance matrix in Fourier space as:
\begin{equation}\label{eq:covariance_forecast}
    \left\langle  \tilde{\delta}(\mathbf{k})\tilde{\delta}^*(\mathbf{k}')   \right\rangle = (2\pi)^3 \delta^D(\mathbf{k}-\mathbf{k}') \left( P_F(\mathbf{k}) + P_w^\perp P^\mathrm{1D}_F(k_\parallel) + P^\mathrm{eff}_N \right) ,
\end{equation}
where $ P_F(\mathbf{k})$ is the 3-dimensional \lya power spectrum (P3D), $P^\mathrm{1D}_F(k_\parallel)$ is the 1-dimensional \lya power spectrum (P1D). $P_w^\perp$ is the power spectrum of the weight overdensity ($\delta_w(\mathbf{x}_\perp) = w(\mathbf{x}_\perp)/\overline{w}-1$), where $w(\mathbf{x}_\perp)$ is the weight assigned to each line of sight. The weight field is treated as white noise in the transverse direction, reflecting the assumption that weights are uncorrelated between different forests. In contrast, within each line of sight, the weights are assumed to be perfectly correlated across pixels. Finally, $P^\mathrm{eff}_N$ is the weighted noise power, dependent on the target density and spectral noise. A detailed description on the estimation of each of the terms in \Cref{eq:covariance_forecast} can be found in \cite{McDonaldForecast}, here we will just give a brief summary. 

In general, $P_w^\perp$ depends on the distribution of galaxies with magnitude and redshift, the survey area, target density and forest length. The weighting also depends on pixel variance $\sigma_N$, which is also a function of magnitude, redshift and observed wavelength (position within the forest)\footnote{Since we are assuming spectral noise is the same everywhere within a forest, we evaluate relevant terms at a mean forest wavelength, rather than iterating across all pixels.}. The noise term $P_N^\mathrm{eff}$ also depends on all of the parameters mentioned above, but features $\sigma_N$ explicitly rather than just having dependence in the weights ($P_N^\mathrm{eff}\propto \langle \sigma_N^2 \rangle)$.

We adopt a SNR of 0.33 \AA$^{-1}$ for the Ly$\alpha$(LBG)$\times$Ly$\alpha$(LBG), Ly$\alpha$(LBG)$\times$LBG, and Ly$\alpha$(LBG)$\times$Ly$\alpha$(LBG)+Ly$\alpha$(LBG)$\times$LBG correlation functions estimates. This corresponds to the median SNR value found in the mock LBG spectra used in our mock-based forecast. We note that this approximation is used because we lack noise estimates as a function of redshift, magnitude, and observed wavelength.

\begin{figure}[tbp]
    \centering
    \includegraphics[width=\textwidth]{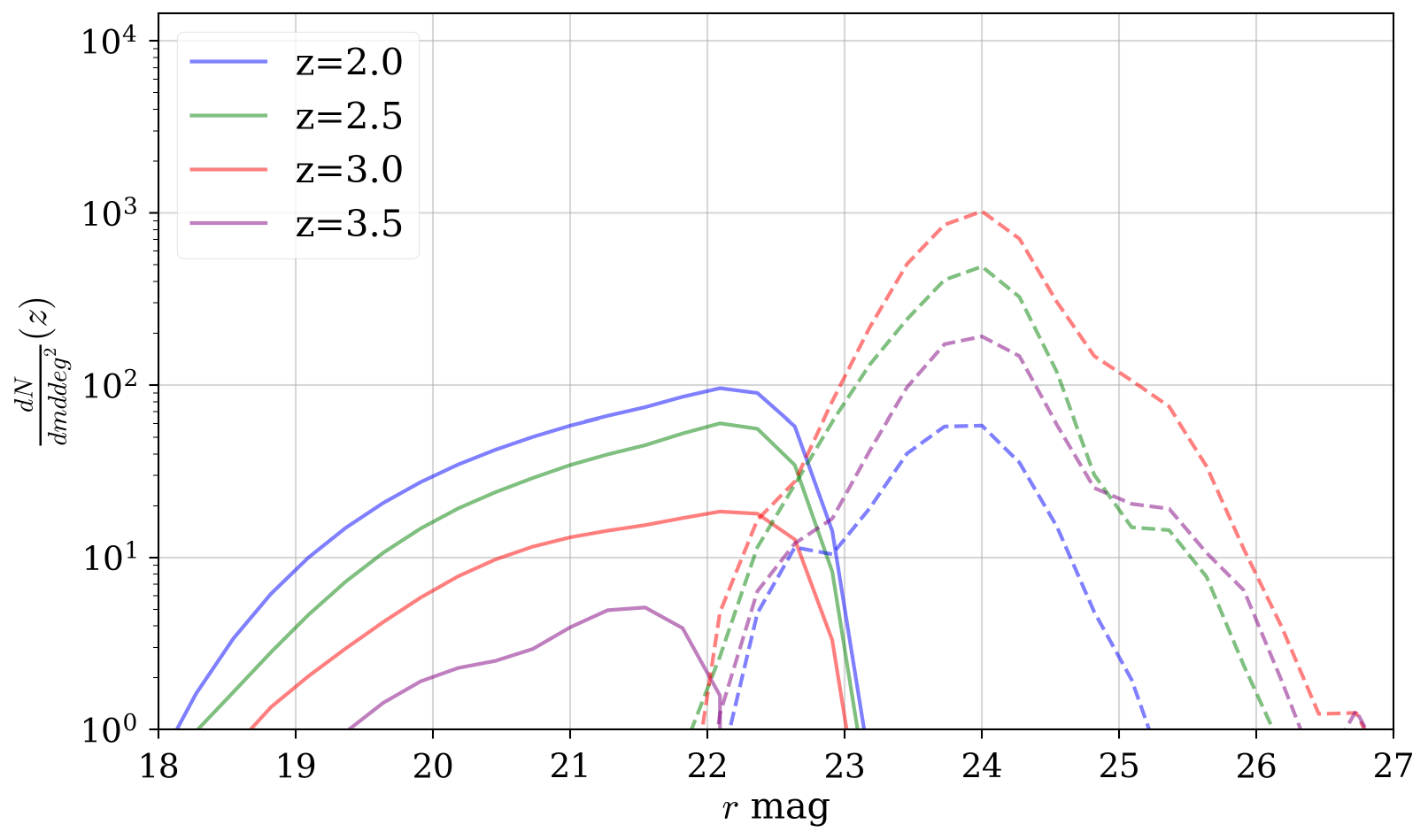}
    \caption{Density distributions, as a function of $r$-band magnitude and redshift, used for both our mock-based and Fisher forecasts. Solid lines: Quasar distribution for the completed DESI survey, with a mean target density of 60\,$\deg^{-2}$. Dashed lines: LBG distribution for a COSMOS-like survey, scaled to a target density 1000\,$\deg^{-2}$.}
    \label{fig:dn_dz_lbg_qso}
\end{figure}

Taking advantage of the flexibility of the Fisher matrix formalism, we also forecast the uncertainties for the LBG$\times$LBG and QSO$\times$QSO auto-correlation as well as for the Ly$\alpha$(QSO)$\times$LBG cross-correlation that combines the DESI quasar forest with the LBG distribution from our hypothetical survey. For all forecasts involving quasars, we adopt the SNR distribution expected for the final DESI survey~\citep{DESI2016a.Science}. The luminosity functions used in these forecasts are shown in \Cref{fig:dn_dz_lbg_qso}, with solid lines denoting quasars and dashed lines representing LBGs, derived from DESI targeting projections and from our COSMOS sample, respectively.

The Fisher forecast results are presented in the second part of \Cref{tab:forecasts}. The two forecasting methods yield consistent results, with the LBG forest auto and cross-correlations combination agreeing within $\lesssim5$\%. The Fisher analysis also shows that the Ly$\alpha$(QSO)$\times$LBG cross-correlation is less constraining than the corresponding Ly$\alpha$(LBG)$\times$LBG measurement. In agreement with the mock spectra forecasts, quasar-based analyses exhibit larger forecasted uncertainties due to their much lower tracer density, despite the higher SNR of quasars. The dense sampling of LBGs compensates for their lower SNR, yielding overall tighter BAO constraints. Since quasar forest analysis typically probe lower redshifts (e.g., $z_{\rm eff}=2.33$ for \DESIDRIILya) than our LBG forest analysis ($z_{\rm eff}=2.7$), the two tracers probe complementary epochs of cosmic evolution. Their joint use therefore enables a continuous mapping of large-scale structure across the range $2 \lesssim z \lesssim 3.5$.

In the transverse direction, the LBG auto-correlation provides the most constraining measurement, owing to the very high LBG target density. We find an average SNR per mode for the LBG auto-correlation of $nP\gtrsim1$, approaching the regime where cosmic variance becomes significant. In the very low shot-noise limit ($nP\gg1$), the benefit of combining tracers for BAO constraints is minimal, and only increasing the survey volume improves precision (see \cite{FontForecast} for discussion). When combining measurements for the results in \Cref{tab:forecasts}, we therefore adopt the high-noise approximation in which parameter errors are combined through a simple inverse-sum. For the Ly$\alpha$(LBG)$\times$LBG and Ly$\alpha$(LBG)$\times$Ly$\alpha$(LBG) correlations, $nP\sim0.3$, meaning we may slightly underestimate the total errors. A more accurate treatment would use the full Fisher matrix including cross-covariances between measurements, which we leave for future work.

\begin{table}[tbp]
\centering
\caption{Forecasted BAO scale uncertainties for a 5,000 $\deg^2$ survey with 1,000~LBGs~deg$^{-2}$ and a mean SNR of 0.33~\AA$^{-1}$, using two approaches: synthetic mock-based fits and Fisher matrix formalism. For comparison, we include analogous forecasts for a quasar forest analysis on a 5,000 $\deg^2$ portion of the completed DESI \lya\ survey. The Fisher forecasts additionally include the LBG and QSO auto-correlations as well as the Ly$\alpha$(QSO)$\times$LBG cross-correlation between the forest from DESI quasars in a completely overlapping area with the LBGs in the hypothetical survey. The final column ($\sigma^{\mathrm{FS}}_{\aAP}$) lists Full Shape (FS) constraints on the Alcock–Paczynski parameter from mock-based forecasts at scales $r > 30~\hMpc$.}\label{tab:forecasts}
\resizebox{\textwidth}{!}{%
\begin{tabular}{lccccc}
Correlation & $\sigma_{\ap}$ (\%) & $\sigma_{\at}$ (\%) & $\sigma_{\aiso}$ (\%) & $\sigma_{\aAP}$ (\%) & $\sigma^{\mathrm{FS}}_{\aAP}$ (\%) \\
\toprule
\multicolumn{6}{l}{\textbf{Mock-based Forecast} \rule{0pt}{4ex} }\\
Ly$\alpha$(LBG)$\times$Ly$\alpha$(LBG)   & 1.08 & 1.13 & 0.57 & 1.90 & 0.67 \\
Ly$\alpha$(LBG)$\times$LBG   & 1.10 & 0.95 & 0.55 & 1.77 & 1.00 \\
Ly$\alpha$(LBG)$\times$Ly$\alpha$(LBG)+Ly$\alpha$(LBG)$\times$LBG  & 0.83 & 0.72 & 0.42 & 1.29 & 0.57 \\\hdashline
Ly$\alpha$(QSO)$\times$Ly$\alpha$(QSO)  & 1.82 & 2.03 & 0.98 & 3.31 & 1.07 \\
Ly$\alpha$(QSO)$\times$QSO  & 2.50 & 2.04 & 1.23 & 3.91 & 2.07 \\
Ly$\alpha$(QSO)$\times$Ly$\alpha$(QSO)+Ly$\alpha$(QSO)$\times$QSO   & 1.63 & 1.54 & 0.83 & 2.73 & 0.94 \\
\hline
\multicolumn{6}{l}{\textbf{Fisher Forecast} \rule{0pt}{4ex} } \\
Ly$\alpha$(LBG)$\times$Ly$\alpha$(LBG)   & 1.11 & 1.00 & 0.57 & 1.80 & -- \\
Ly$\alpha$(LBG)$\times$LBG   & 1.26 & 0.86 & 0.62 & 1.82 & -- \\
Ly$\alpha$(LBG)$\times$Ly$\alpha$(LBG)+Ly$\alpha$(LBG)$\times$LBG   & 0.87 & 0.68 & 0.44 & 1.34 & -- \\
LBG$\times$LBG & 1.04 & 0.59 & 0.51 & 1.41 & -- \\\hdashline
Ly$\alpha$(QSO)$\times$LBG  & 1.78 & 1.38 & 0.89 & 2.72 & -- \\ 
Ly$\alpha$(QSO)$\times$Ly$\alpha$(QSO)  & 1.89 & 1.99 & 0.99 & 3.34 & -- \\
Ly$\alpha$(QSO)$\times$QSO  & 2.91 & 2.35 & 1.43 & 4.53 & -- \\
Ly$\alpha$(QSO)$\times$Ly$\alpha$(QSO)+Ly$\alpha$(QSO)$\times$QSO   & 1.15 & 1.08 & 0.58 & 1.92 & -- \\
QSO$\times$QSO & 4.00 & 2.58 & 1.94 & 5.70 & -- \\

\end{tabular}}
\end{table}

\section{Summary and Conclusions}\label{sec:summary_conclusions}
Measurements of the \lya\ forest from quasar spectra have delivered groundbreaking cosmological constraints at $z>2$, including a sub-percent precision BAO measurement at $z_{\rm eff}=2.33$ from \DESIDRIILya. However, such studies remain limited by the relatively low number density of quasars. In this work, we present the first 3D \lya\ forest correlation analysis based exclusively on Lyman-break galaxy (LBG) spectra, using a pioneering DESI pilot sample in the COSMOS and XMM fields selected via broadband imaging~\citep{Ruhlmann-Kleider:2024LBG,Payerne:2024LBGs}. By adapting the DESI \lyaf\ analysis pipeline, we demonstrate that it performs robustly even on lower signal-to-noise ratio (SNR) LBG spectra.

We measure the \lya\ forest auto-correlation function on a $0 < \rp,\rt < 48\ \hMpc$ grid at an effective redshift $z_{\rm eff}=2.70$, based on 4,151 forests extracted exclusively from LBG spectra. Remarkably, the results show excellent agreement with those derived from the significantly higher-SNR quasar sample in \DESIDRIILya, demonstrating that LBG forests, even at the relatively low SNR of our sample, can reliably trace the intergalactic medium across large volumes. In addition to the auto-correlation, we analyze the Ly$\alpha$–galaxy cross-correlation using the positions of 4,931 LBGs and 8,431 Lyman-$\alpha$ emitters (LAEs) in the same fields. The Ly$\alpha$–galaxy cross-correlation provides a powerful tool for calibrating galaxy redshifts, allowing us to correct systematic redshift offsets of $\Delta v_{\rm LBG} = -169 \pm 22\ \kms$ for LBGs and $\Delta v_{\rm LAE} = -241 \pm 20\ \kms$ for LAEs.

An important distinction between LBG and quasar-based analyses is the significantly higher rate of catastrophic redshift outliers and the lower sample purity in LBG samples. For our DESI LBG dataset, we estimate an outlier fraction of $\sim$10\% for galaxies with CNN confidence level $\mathrm{CL}>0.995$, consistent with the findings of \RuhlmannLBG, who additionally report a $\sim$94\% purity at the same threshold. Such contamination directly suppresses the observed clustering amplitude and increases the uncertainty of correlation measurements. In contrast, DESI \lya\ quasars typically exhibit catastrophic redshift fractions of $\lesssim 2\%$ and purities exceeding 98\%~\citep{Alexander:2022pql,RedrockQSO.Brodzeller.2023}. The relatively larger uncertainty in LBG samples highlights the need for refined mitigation and modeling strategies to fully exploit the statistical power of future dense LBG surveys, alongside a systematic characterization of effects impacting the \lyaf\ specifically. Future improvements may build upon the experience from DESI galaxy clustering analyses, which employ redshift-failure weighting schemes and mock-based evaluations of catastrophic redshift impacts, along with additional systematic mitigation strategies~\citep{KP3s3-Krolewski,KP3s4-Yu}, as well as from \lya\ correlation studies that address correlated noise~\citep{KP6s5-Guy} and other potential systematics that could specifically affect LBG-based forest measurements.

Despite these challenges, our results demonstrate that the \lya\ forest extracted from LBG spectra in DESI-like surveys can serve as a complementary cosmological probe to quasar-based analyses and establish a promising path toward mapping large-scale structure at redshifts $z \gtrsim 2$. Looking ahead, we forecast the BAO measurement capabilities of a future spectroscopic survey targeting 1,000 LBGs per square degree over a 5,000 deg$^2$ footprint, assuming a median SNR of $\sim 0.33$~\AA$^{-1}$. Using both Fisher-matrix and mock-spectra forecasts independently, we find that combining the \lya\ forest auto-correlation and the \lyaxlbg\ cross-correlation can deliver competitive constraints at $z>2.6$, with expected uncertainties of $\sigma_{\ap}=0.8\%$ and $\sigma_{\at}=0.7\%$, corresponding to $\sigma_{\aiso}=0.4\%$ in the isotropic and $\sigma_{\aAP}=1.3\%$ in the Alcock–Paczynski (AP) directions. This constraining power is further enhanced when the BAO analysis is complemented by a full-shape analysis of the \lya\ forest, yielding a predicted uncertainty of $\sigma_{\aAP} = 0.6\%$.

This precision is crucial in light of recent DESI results hinting at possible deviations from the standard dark energy model~\citep{DESI2024.VI.KP7A,DESI.DR2.BAO.cosmo}. Achieving such accuracy at high redshift will require improved sample purity, optimized target selection, deeper imaging, and a careful characterization of systematics, as highlighted by our analysis. This level of performance is within reach of next-generation spectroscopic programs such as DESI-II~\citep{DESI-II:2022}, MUST~\citep{Zhao:2024MUST}, SPEC-S5~\citep{Spec-S5:2025uom}, and WST~\citep{WST:2024zvm}, especially when combined with wide-area imaging from the DESI Legacy Surveys~\citep{LS.Overview.Dey.2019}, UNIONS~\citep{2025UNIONS}, Euclid~\citep{Euclid:2021icp}, and LSST~\citep{LSSTDarkEnergyScience:2012kar}. As such, the \lya\ forest from LBGs represents a key component of next-generation high-redshift cosmological probes.

\section*{Data Availability}
All data points presented in the figures of this article will be made publicly available upon publication at \url{https://doi.org/10.5281/zenodo.17242512}, in accordance with the DESI data management policy. The DESI Early Data Release and Data Release 1 are publicly accessible at: \url{https://data.desi.lbl.gov/doc/releases/}.

\section*{Acknowledgements}
We are grateful to Pat McDonald for developing the original \texttt{C++} version of the Ly$\alpha$ fisher forecast code, which served as the basis for the \texttt{lyaforecast} code.

HKHA, EA, and CY acknowledge support from the French National Research Agency (ANR) under grant ANR-22-CE31-0009 (HZ-3D-MAP project) and grant ANR-22-CE92-0037 (DESI-Lya project).

CG, LC and AFR acknowledge support from the European Union’s Horizon Europe research and innovation programme (COSMO-LYA, grant agreement 101044612). CG is also partially supported by the Spanish Ministry of Science and Innovation (MICINN) under grants PGC-2018-094773-B-C31 and SEV-2016-0588. AFR acknowledges financial support from the Spanish Ministry of Science and Innovation under the Ramon y Cajal program (RYC-2018-025210) and the PGC2021-123012NB-C41 project. IFAE is partially funded by the CERCA program of the Generalitat de Catalunya. 

This material is based upon work supported by the U.S. Department of Energy (DOE), Office of Science, Office of High-Energy Physics, under Contract No. DE–AC02–05CH11231, and by the National Energy Research Scientific Computing Center, a DOE Office of Science User Facility under the same contract. Additional support for DESI was provided by the U.S. National Science Foundation (NSF), Division of Astronomical Sciences under Contract No. AST-0950945 to the NSF’s National Optical-Infrared Astronomy Research Laboratory; the Science and Technology Facilities Council of the United Kingdom; the Gordon and Betty Moore Foundation; the Heising-Simons Foundation; the French Alternative Energies and Atomic Energy Commission (CEA); the National Council of Humanities, Science and Technology of Mexico (CONAHCYT); the Ministry of Science, Innovation and Universities of Spain (MICIU/AEI/10.13039/501100011033), and by the DESI Member Institutions: \url{https://www.desi.lbl.gov/collaborating-institutions}. Any opinions, findings, and conclusions or recommendations expressed in this material are those of the author(s) and do not necessarily reflect the views of the U. S. National Science Foundation, the U. S. Department of Energy, or any of the listed funding agencies.

The authors are honored to be permitted to conduct scientific research on I'oligam Du'ag (Kitt Peak), a mountain with particular significance to the Tohono O’odham Nation.

\bibliographystyle{JHEP.bst}
\bibliography{references,DESI_supporting_papers}

\appendix 
\counterwithin{figure}{section}
\counterwithin{table}{section}
\section{Correction for the Bias in Redshift Estimate}\label{appendix:redshift_correction}
The automatic redshift estimates for our LBG and LAE samples were calibrated against visually inspected spectra, which primarily rely on strong absorption features from low-ionization lines (\ion{Si}{II}{1260}, \ion{O}{I}{1302}+ \ion{Si}{II}{1304}, \ion{C}{II}{1335}) and high-ionization lines (the \ion{Si}{IV}{1394}, \ion{Si}{IV}{1403} doublet). As shown by \cite{Shapley:2003LBG}, these interstellar absorption lines are systematically blueshifted relative to the systemic redshift of the galaxy, leading to an underestimation of the true redshift in visual inspections.

In our correlation-function framework (see \Cref{sec:analysis}), a systematic error in the galaxy redshift translates into a shift of the cross-correlation along the line-of-sight. By measuring the \lyaxlbg cross-correlation using automatic redshift estimates we find a $\Delta r_{\rm \parallel, LBG} = -1.54\pm 0.20\ \hMpc$ offset. Similarly, for \lyaxlae we measure an offset of $\Delta r_{\rm \parallel, LAE} = -2.20\pm 0.18\ \hMpc$. These displacements correspond to a velocity offsets of $\Delta v_{\rm LBG} = -169 \pm 22\ \kms$ and $\Delta v_{\rm LAE} = -241 \pm 20\ \kms$ at our effective redshift, respectively. 

The LBG velocity offset we measure agrees with the typical $-100$ to $-200\ \kms$ blueshifts of interstellar absorption lines observed in $z \sim 2-3$ star-forming galaxies~\citep[e.g.,][]{Pettini2001,Shapley:2003LBG,Steidel2010,Rakic2011}. For LAEs, the inferred offset is in line with values reported for interstellar absorption features in LAE spectra~\citep[e.g.,][]{Hashimoto2013,Shibuya2014}, which are expected to be comparable to those of LBGs.

\begin{figure}
    \centering
    \includegraphics[width=\textwidth]{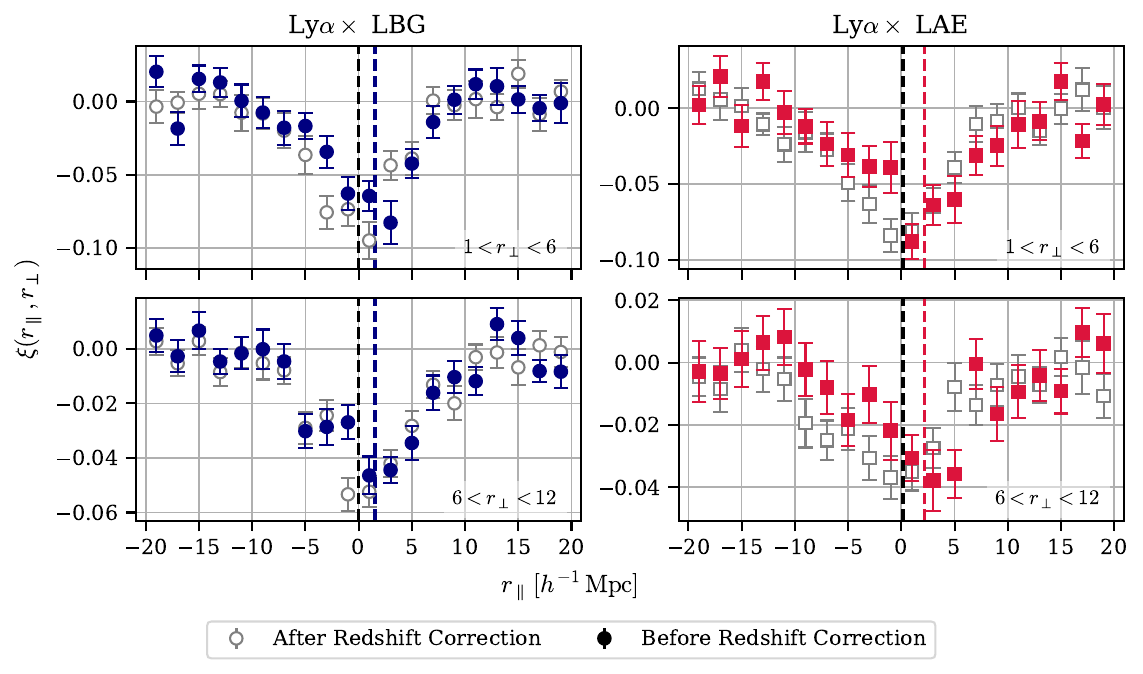}
\caption{Comparison of the \lyaxlbg (left) and \lyaxlae (right) cross-correlation signals before (gray open symbols) and after (colored filled symbols) applying the redshift-offset correction. The measurements are shown as a function of transverse separation $\rp$, averaged over two line-of-sight separation bins: $1 < \rt < 6\ \hMpc$ (top panels) and $6 < \rt < 12\ \hMpc$ (bottom panels). For improved visibility, we restrict the $\rp$ range to $-20 < \rp < 20\ \hMpc$. Vertical dashed lines indicate the best-fit $\Delta r_{\parallel, \rm X}$ offsets for each tracer, with colored and black lines corresponding to the results before and after correcting redshift estimates, respectively.}
    \label{fig:cross_zshift}
\end{figure}

We correct for these redshift offsets by applying a shift to each galaxy redshift, using the transformation $z' \rightarrow z - \frac{\Delta v_{\rm X}}{c}(1+z)$, where $c$ is the speed of light. The inferred residual offsets after these corrections are $\Delta r_{\rm \parallel, LBG} = 0.04\pm0.18\ \hMpc$ and $\Delta r_{\rm \parallel, LAE} = -0.13\pm0.17\ \hMpc$, both consistent with zero which indicates that the procedure effectively removes the systematic bias. 

To illustrate the impact of this correction, \Cref{fig:cross_zshift} compares the \lyaxlbg (left panels) and \lyaxlae (right panels) cross-correlation functions before and after applying the redshift adjustments. The correlations are shown as a function of transverse separation $\rp$, averaged over line-of-sight bins of $1 < \rt < 6\ \hMpc$ (top panels) and $6 < \rt < 12\ \hMpc$ (bottom panels). Vertical dashed lines mark the measured offsets $\Delta r_{\parallel, \rm X}$ for each tracer, with colored lines indicating the best-fit offset values prior redshift estimate corrections and black lines the results after corrections are applied.

\section{Impact of Sample Purity on Posterior Distributions}\label{appendix:posteriors}
\begin{figure}
    \centering
    \includegraphics[width=\textwidth]{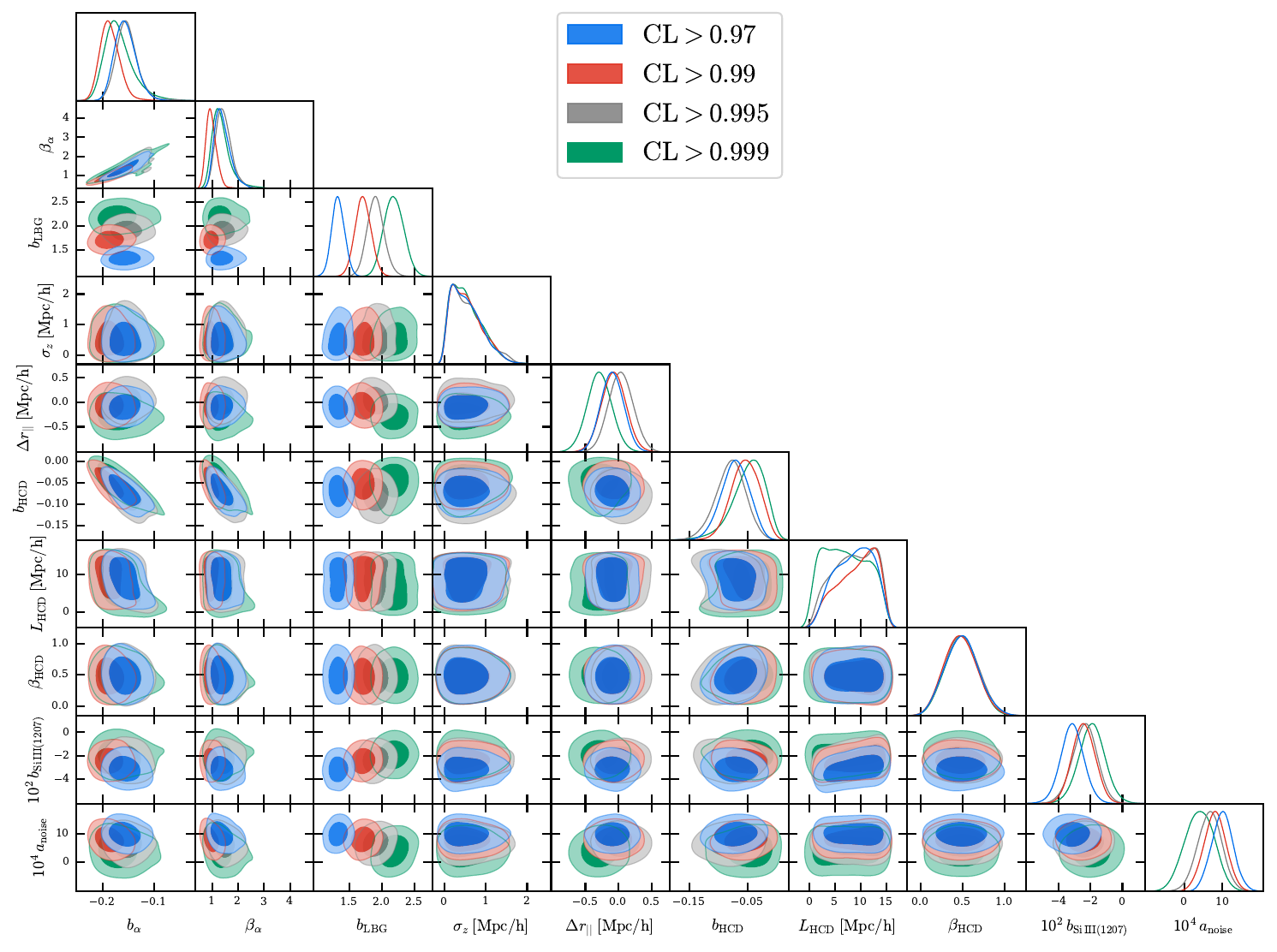}
    \caption{Posterior distributions for every model parameter (see \Cref{sec:model}), from the \combinedLBG joint-fit of the combined COSMOS+XMM LBG sample fit over $1<r<28\ \hMpc$ at $z_{\rm eff} \sim 2.70$, for different CNN confidence cuts.}
    \label{fig:full_posteriors}
\end{figure}

In \Cref{subsec:param_contraints}, we presented key constraints on the model parameters derived from the combined fits of the \lyaxlya and \lyaxlbg correlation functions. Here, we expand on that summary by showing the full posterior distributions for all free parameters involved in the joint auto and cross-correlation fit over the range $1 < r < 28\ \hMpc$. We also examine the impact of the CNN confidence level (CL) threshold on the resulting parameter estimates. These distributions provide a more complete view of parameter degeneracies and the sensitivity of the results to sample purity. The full triangle plot for \combinedLBG\ is shown in \Cref{fig:full_posteriors}.

We begin by examining metal and noise contamination. In our baseline fit, which adopts $\mathrm{CL} > 0.995$, we measure a \ion{Si}{III}{1207} linear bias of $b_{\rm \ion{Si}{III}{1207}} = (-2.36 \pm 0.77) \times 10^{-2}$ and detect correlated noise with an amplitude of $a_{\rm noise} = (6.0 \pm 3.2) \times 10^{-4}$. These values are significantly larger than those obtained from the joint auto and cross-correlation fit of \DESIDRIILya\ quasars, where we measured $b_{\rm \ion{Si}{III}{1207}} = (-0.97 \pm 0.11) \times 10^{-2}$ and $a_{\rm noise} = (3.39 \pm 0.31) \times 10^{-4}$. Note that in the quasar-based results, forest–forest and forest–quasar pairs were restricted in the correlation-function estimation, and DLAs were not masked, to match the effective redshift and observational conditions of our LBG sample. One might therefore have expected closer agreement with the LBG-based measurements.

A plausible explanation for this discrepancy lies in differences in sample purity: as shown in \Cref{fig:full_posteriors}, both $b_{\rm \ion{Si}{III}{1207}}$ and $a_{\rm noise}$ systematically decrease with increasing CL. This trend suggests that the fitted parameters are sensitive to contamination from misclassified objects or interlopers. Higher CL thresholds (i.e., higher purity) appear to suppress these effects and may help bring the LBG-derived values into closer agreement with those from the quasar-based analysis. For correlated noise, an additional contributing factor may be the lower SNR of the LBG spectra at lower CL thresholds, which can amplify the apparent noise signature in the auto-correlation.

Next, we examine the LBG, \lya, and HCD parameters. As discussed in \cref{subsec:param_contraints}, we find that the inferred LBG bias increases with CL threshold, while the LBG redshift error parameters remain relatively stable. This behavior suggests a significant fraction of interlopers in the LBG sample at lower CLs, which biases the inferred clustering amplitude. A similar effect is expected for the \lya\ linear bias. However, as shown in \Cref{fig:full_posteriors}, $b_\alpha$ is degenerate with both the \lya\ RSD parameter and the HCD linear bias, complicating the interpretation and limiting our ability to isolate the purity dependence.

In our baseline analysis, we obtain an HCD linear bias of $b_{\rm HCD} = -0.079^{+0.027}_{-0.024}$. For $\beta_{\rm HCD}$, we recover values consistent with the Gaussian prior set to $0.5 \pm 0.20$, while we are not able to robustly constrain $L_{\rm HCD}$. We expect this limitation to be alleviated as the LBG sample size increases and additional constraints (e.g., masking of DLAs) are applied.

As a complementary test to those presented in \Cref{tab:fit_variations}, we also examined the trend in $b_\alpha$ as a function of CL threshold while fixing $\beta_\alpha$ and the HCD parameters to the values derived from \DESIDRIILya\ quasars at the effective redshift of our sample, with DLAs unmasked to match the LBG conditions. We find a consistent decrease in $b_\alpha$ with increasing sample purity: $b_\alpha = -0.1276 \pm 0.0033$ for $\mathrm{CL} > 0.97$, $-0.1286 \pm 0.0035$ for $\mathrm{CL} > 0.99$, $-0.1304 \pm 0.0035$ for $\mathrm{CL} > 0.995$, and $-0.1339 \pm 0.0039$ for $\mathrm{CL} > 0.999$. This trend is similar as the observed in the LBG linear bias and reinforces the interpretation that sample purity is having a considerable impact on the inferred clustering amplitudes of both the \lya\ forest and LBGs.


\section{Author Affiliations}
\label{sec:affiliations}

\noindent \hangindent=.5cm $^{1}${Institut d'Astrophysique de Paris. 98 bis boulevard Arago. 75014 Paris, France}

\noindent \hangindent=.5cm $^{2}${IRFU, CEA, Universit\'{e} Paris-Saclay, F-91191 Gif-sur-Yvette, France}

\noindent \hangindent=.5cm $^{3}${Department of Physics \& Astronomy, University College London, Gower Street, London, WC1E 6BT, UK}

\noindent \hangindent=.5cm $^{4}${Institut de F\'{i}sica d’Altes Energies (IFAE), The Barcelona Institute of Science and Technology, Edifici Cn, Campus UAB, 08193, Bellaterra (Barcelona), Spain}

\noindent \hangindent=.5cm $^{5}${Universit\'{e} Clermont-Auvergne, CNRS, LPCA, 63000 Clermont-Ferrand, France}

\noindent \hangindent=.5cm $^{6}${Lawrence Berkeley National Laboratory, 1 Cyclotron Road, Berkeley, CA 94720, USA}

\noindent \hangindent=.5cm $^{7}${Department of Physics, Boston University, 590 Commonwealth Avenue, Boston, MA 02215 USA}

\noindent \hangindent=.5cm $^{8}${Department of Physics and Astronomy, The University of Utah, 115 South 1400 East, Salt Lake City, UT 84112, USA}

\noindent \hangindent=.5cm $^{9}${Instituto de F\'{\i}sica, Universidad Nacional Aut\'{o}noma de M\'{e}xico,  Circuito de la Investigaci\'{o}n Cient\'{\i}fica, Ciudad Universitaria, Cd. de M\'{e}xico  C.~P.~04510,  M\'{e}xico}

\noindent \hangindent=.5cm $^{10}${NSF NOIRLab, 950 N. Cherry Ave., Tucson, AZ 85719, USA}

\noindent \hangindent=.5cm $^{11}${University of California, Berkeley, 110 Sproul Hall \#5800 Berkeley, CA 94720, USA}

\noindent \hangindent=.5cm $^{12}${Departamento de F\'isica, Universidad de los Andes, Cra. 1 No. 18A-10, Edificio Ip, CP 111711, Bogot\'a, Colombia}

\noindent \hangindent=.5cm $^{13}${Observatorio Astron\'omico, Universidad de los Andes, Cra. 1 No. 18A-10, Edificio H, CP 111711 Bogot\'a, Colombia}

\noindent \hangindent=.5cm $^{14}${Institut d'Estudis Espacials de Catalunya (IEEC), c/ Esteve Terradas 1, Edifici RDIT, Campus PMT-UPC, 08860 Castelldefels, Spain}

\noindent \hangindent=.5cm $^{15}${Institute of Cosmology and Gravitation, University of Portsmouth, Dennis Sciama Building, Portsmouth, PO1 3FX, UK}

\noindent \hangindent=.5cm $^{16}${Institute of Space Sciences, ICE-CSIC, Campus UAB, Carrer de Can Magrans s/n, 08913 Bellaterra, Barcelona, Spain}

\noindent \hangindent=.5cm $^{17}${University of Virginia, Department of Astronomy, Charlottesville, VA 22904, USA}

\noindent \hangindent=.5cm $^{18}${Departamento de F\'{\i}sica, DCI-Campus Le\'{o}n, Universidad de Guanajuato, Loma de
l Bosque 103, Le\'{o}n, Guanajuato C.~P.~37150, M\'{e}xico}

\noindent \hangindent=.5cm $^{19}${Fermi National Accelerator Laboratory, PO Box 500, Batavia, IL 60510, USA}

\noindent \hangindent=.5cm $^{20}${Steward Observatory, University of Arizona, 933 N. Cherry Avenue, Tucson, AZ 85721,
USA}

\noindent \hangindent=.5cm $^{21}${Department of Physics and Astronomy, University of California, Irvine, 92697, USA}

\noindent \hangindent=.5cm $^{22}${Sorbonne Universit\'{e}, CNRS/IN2P3, Laboratoire de Physique Nucl\'{e}aire et de Hau
tes Energies (LPNHE), FR-75005 Paris, France}

\noindent \hangindent=.5cm $^{23}${Departament de F\'{i}sica, Serra H\'{u}nter, Universitat Aut\`{o}noma de Barcelona,
08193 Bellaterra (Barcelona), Spain}

\noindent \hangindent=.5cm $^{24}${Center for Cosmology and AstroParticle Physics, The Ohio State University, 191 West
Woodruff Avenue, Columbus, OH 43210, USA}

\noindent \hangindent=.5cm $^{25}${Department of Astronomy, The Ohio State University, 4055 McPherson Laboratory, 140 W
 18th Avenue, Columbus, OH 43210, USA}

\noindent \hangindent=.5cm $^{26}${The Ohio State University, Columbus, 43210 OH, USA}

\noindent \hangindent=.5cm $^{27}${Instituci\'{o} Catalana de Recerca i Estudis Avan\c{c}ats, Passeig de Llu\'{\i}s Com
panys, 23, 08010 Barcelona, Spain}

\noindent \hangindent=.5cm $^{28}${Department of Physics and Astronomy, University of Waterloo, 200 University Ave W, W
aterloo, ON N2L 3G1, Canada}

\noindent \hangindent=.5cm $^{29}${Perimeter Institute for Theoretical Physics, 31 Caroline St. North, Waterloo, ON N2L
 2Y5, Canada}

\noindent \hangindent=.5cm $^{30}${Waterloo Centre for Astrophysics, University of Waterloo, 200 University Ave W, Wate
rloo, ON N2L 3G1, Canada}

\noindent \hangindent=.5cm $^{31}${Instituto de Astrof\'{i}sica de Andaluc\'{i}a (CSIC), Glorieta de la Astronom\'{i}a,
 s/n, E-18008 Granada, Spain}

\noindent \hangindent=.5cm $^{32}${Departament de F\'isica, EEBE, Universitat Polit\`ecnica de Catalunya, c/Eduard Mari
stany 10, 08930 Barcelona, Spain}

\noindent \hangindent=.5cm $^{33}${Department of Physics and Astronomy, Sejong University, 209 Neungdong-ro, Gwangjin-g
u, Seoul 05006, Republic of Korea}

\noindent \hangindent=.5cm $^{34}${CIEMAT, Avenida Complutense 40, E-28040 Madrid, Spain}

\noindent \hangindent=.5cm $^{35}${Department of Physics, University of Michigan, 450 Church Street, Ann Arbor, MI 4810
9, USA}

\noindent \hangindent=.5cm $^{36}${University of Michigan, 500 S. State Street, Ann Arbor, MI 48109, USA}

\noindent \hangindent=.5cm $^{37}${National Astronomical Observatories, Chinese Academy of Sciences, A20 Datun Road, Ch
aoyang District, Beijing, 100101, P.~R.~China}
\end{document}